\documentclass[a4paper,11pt]{article}
\pdfoutput=1
\usepackage{jcappub}

\usepackage{tikz,xcolor,hyperref}
\usepackage{amsmath, amssymb, amsthm, graphicx, epsfig, fancyhdr,epsfig, slashed}
\usepackage{tikzsymbols}
\usepackage{natbib}
\usepackage{float}

\newcommand{\baz}{\begin{array}{cc}}
\newcommand{\ux}{U\left(1\right)_X}

\newcommand{\lhs}{\lambda_{HS}}

\newcommand{\lh}{\lambda_H}

\def\be{\begin{equation}}
\def\ee{\end{equation}}
\def\beq{\begin{equation}\begin{aligned}}
\def\eeq{\end{aligned}\end{equation}}

\begin{document}

\title{Probing pre-BBN era with Scale Invariant FIMP}


\author[a]{Basabendu Barman\,, 
}
\author[b]{ Anish Ghoshal
}
\affiliation[a]{\,Centro de Investigaciones, Universidad Antonio Nari\~{n}o\\Carrera 3 este \# 47A-15, Bogot{\'a}, Colombia}
\affiliation[b]{Institute of Theoretical Physics, Faculty of Physics, University of Warsaw,\\ ul.  Pasteura 5, 02-093 Warsaw, Poland}
\emailAdd{basabendu88barman@gmail.com}
\emailAdd{anish.ghoshal@fuw.edu.pl}

\abstract{Detecting dark matter (DM) relic via freeze-in is difficult in laboratories due to smallness of the couplings involved. However, a non-standard cosmological history of the Universe, prior to Big Bang Nucleosynthesis (BBN), can dramatically change this scenario. In this context, we study the freeze-in production of dark matter (DM) in classically scale invariant $U(1)_X$ gauge extension of the Standard Model (SM), recently dubbed as the \textit{Scale Invariant FIMP Miracle}. We assume an additional species dominates the energy density of the Universe at early times, causing the expansion rate at a given temperature to be larger than that in the standard radiation-dominated case. We find, the \textit{out-of-equilibrium} scattering processes involving particles in the thermal bath lead to significantly suppressed DM production in this era, thereby enhancing the couplings between the visible and the dark sector (by several orders of magnitude) to satisfy the observed DM abundance, and improving the detection prospects for freeze-in in turn. Scale invariance of the underlying theory leaves only four free parameters in the model: the DM mass $m_X$, the gauge coupling $g_X$, the temperature of transition $T_R$ from early scalar-dominated to radiation-dominated era and the power-law dependence $n$ of this temperature. We show, within this minimal set-up, experiments like FASER, MATHUSLA, DUNE, SHiP will be probing various cosmological models depending on the choice of $\{n,\,T_R\}$ that also satisfy the PLANCK observed relic density bound. Moreover, due to the presence of a naturally light scalar mediator, the direct detection of the DM at XENON1T, PandaX-4T or XENONnT becomes relevant for Higgs-scalar mixing $\sin\theta\simeq\{10^{-5}-10^{-3}\}$, thus providing complementary probes for freeze-in, as well as for non-standard cosmological pre-BBN era.
}

\begin{flushright}
  PI/UAN-2021-712FT
\end{flushright}
\maketitle
\section{Introduction}
\label{sec:intro}
With the Higgs boson being discovered at the LHC~\cite{ATLAS:2012yve, CMS:2012qbp}, the so-called Higgs mechanism~\cite{PhysRevLett.13.321, PhysRevLett.13.508, Guralnik:1964eu} for generating masses of all Standard Model (SM) particles in a gauge invariant manner was verified but the Higgs itself carries an ad-hoc negative mass term at the electroweak (EW) scale which is unstable against the Planck scale induced diverging quantum corrections~\cite{PhysRevD.20.2619}. Pathways to address this well-known ``naturalness problem" have been the guiding principle behind numerous beyond the SM (BSM) extensions, like supersymmetry (SUSY), Higgs compositeness, extra dimensions, to name a few (for a review, see Ref.~\cite{Giudice:2017pzm}) \footnote{In context to non-local QFT with higher-derivatives, motivated from string field theories, see \cite{Ghoshal:2017egr,Buoninfante:2018gce,Ghoshal:2018gpq,Ghoshal:2020lfd} for solving naturalness.}. Alternatively scale invariance if assumed to be a symmetry of the action at the classical level naturally gets rid of the $\mu^2$ mass term or any other dimensionful parameters \footnote{Scale-invariance, or more generically conformal invariance, has been a direction of model-building for the hierarchy problem in the SM~\cite{Adler:1982ri,Coleman:1973jx,Salvio:2014soa,Einhorn:2014gfa, Ghorbani:2015xvz,Einhorn:2016mws,Einhorn:2015lzy,Foot:2007iy,AlexanderNunneley:2010nw,Englert:2013gz,Hambye:2013sna,Farzinnia:2013pga,Altmannshofer:2014vra,Holthausen:2013ota,Salvio:2014soa,Einhorn:2014gfa,Kannike:2015apa,Farzinnia:2015fka,Kannike:2016bny,Ghoshal:2022hyc, Ghorbani:2022muk}, and in cosmology provides naturally flat inflationary potentials \cite{Khoze:2013uia,Kannike:2014mia,Rinaldi:2014gha,Salvio:2014soa,Kannike:2015apa,Kannike:2015fom,Barrie:2016rnv,Tambalo:2016eqr}, and leads to very strong first-order phase transitions in early Universe and consequently high amplitude detectable gravitational wave (GW) signals~\cite{Jaeckel:2016jlh, Marzola:2017jzl, Iso:2017uuu, Ghorbani:2017lyk, Baldes:2018emh, Prokopec:2018tnq, Brdar:2018num, Marzo:2018nov, Ghoshal:2020vud}. }. In the SM itself as pointed out by Coleman and Weinberg (CW) in their seminal work in Ref.~\cite{PhysRevD.7.1888}, this scale-genesis of electroweak scale works in principle, but given the experimentally observed masses of top quark and electroweak gauge bosons the potential in unstable. However, several BSM scenarios with additional bosonic contributions\footnote{Recently the Neutrino Option idea considers threshold corrections (as an alternative to bosonic corrections) to generate the Higgs mass via fermionic loop~\cite{Brivio:2017dfq}.} to the Higgs work with a ``hidden sector" that via dimensional transmutation leads to generation of the SM Higgs mass~\cite{Foot:2007as, Iso:2009ss, Iso:2009nw, Steele:2012av, Okada:2012sg, Steele:2013fka, Farzinnia:2013pga, Englert:2013gz, Hambye:2013dgv, Kang:2014cia, Steele:2014dsa, Kang:2015aqa, Hambye:2013sna, Kubo:2014ova, Das:2015nwk, Humbert:2015epa, Humbert:2015yva, Haba:2015lka, Haba:2015yfa, Das:2016zue, Hambye:2018qjv,  Kubo:2018kho,Brdar:2018vjq, YaserAyazi:2019caf, Mohamadnejad:2019vzg, Okada:2019opp, Mohapatra:2019ysk,Okada:2020evk, Mohapatra:2020bze, Gialamas:2021enw, Li:2021pnv,Ghorbani:2021rgs,Ghoshal:2022hyc}. 

Such a context to dark sector BSM set-up naturally raises the motivation for solving the dark matter (DM) problem of the Universe. The origin and composition of such a particle DM hypothesis is amongst one of the biggest open questions ~\cite{Jungman:1995df,Bertone:2004pz,Feng:2010gw}. The Weakly Interacting Massive Particles (WIMPs) having a thermal relic abundance, naturally close to the observed DM cosmological abundance, require a DM-SM interaction cross-section of the order ~\cite{Lee:1977ua,Scherrer:1985zt,Srednicki:1988ce} $\langle \sigma_{\rm th} v_{\rm rel} \rangle \simeq 3 \times 10^{-26} \, \text{cm}^3 \, \text{sec}^{-1}$,
where the brackets denote thermally averaged quantity and $v_{\rm rel}$ is the M\o{}ller velocity (see~\cite{Gondolo:1990dk} for details). The cross section $\sigma_{\rm th} v_{\rm rel}$ is typical of the order of SM electroweak interactions, and hence this is often referred to as the ``WIMP miracle.'' With great experimental efforts over the last 30 years we have found no clinching evidence of the existence of the such thermally frozen out DM; in fact we have obtained strong bounds on the interaction strength between DM and the SM particles from direct detection experiments~\cite{PandaX-II:2017hlx,PandaX:2018wtu,XENON:2020kmp,XENON:2018voc,LUX-ZEPLIN:2018poe,DARWIN:2016hyl}, indirect detection~\cite{HESS:2016mib,MAGIC:2016xys} and through collider searches, for example, at the Large Hadron Collider (LHC)~\cite{ATLAS:2017bfj,CMS:2017zts, Kahlhoefer:2017dnp}. 

This provides the motivation to consider alternative non-thermal DM production, particularly the observed DM abundance may have been generated out of equilibrium by the so-called {\it freeze-in} mechanism~\cite{McDonald:2001vt, Hall:2009bx, Bernal:2017kxu}. In this scenario, the DM particle couples to the visible sector very feebly, so that chemical equilibrium is never achieved in the early Universe. The DM particles are mostly produced by the decay or via annihilation of the bath particles, until the production stops due to cooling of the bath temperature below the relevant mass scale of the mediator that connects the DM with the visible sector. Typically, {\it Infra-red (IR)} freeze-in is important at lower temperature~\cite{McDonald:2001vt, Hall:2009bx, Chu:2011be, Bernal:2017kxu, Duch:2017khv, Biswas:2018aib, Heeba:2018wtf, Barman:2019lvm}, while  {\it Ultra-violate (UV)} freeze-in at higher temperatures (roughly the reheating temperature of the Universe)~\cite{Hall:2009bx, Elahi:2014fsa, Chen:2017kvz, Bernal:2019mhf, Biswas:2019iqm, Barman:2020plp, Barman:2020ifq, Bernal:2020bfj, Bernal:2020qyu, Barman:2021tgt}. Due to the smallness of the coupling strengths involved in the processes, it is quite challenging to see any laboratory imprints (direct searches or otherwise) of frozen in DM. One of the plausible pathways to detect freeze-in dark matter is supposing DM production in the early Universe proceeds via the decay of thermal bath particles. This necessarily means that the  feeble couplings associated with such decays would make the dark sector particles very long-lived and be searched at the LHC and beyond (see, for example, Ref.~\cite{Curtin:2018mvb}). Though this provides a very promising and suitable avenue to probe freeze-in DM scenarios~\cite{Hessler:2016kwm,Belanger:2018sti}, but on the other hand, if the DM freeze-in production happens via scattering, then depending on the nature of the portal (scalar-portal, vector-portal, etc.) one may look for the long lived mediator particles in current and upcoming lifetime and intensity frontier experimental searches~\cite{Heeba:2019jho, Mohapatra:2019ysk, Mohapatra:2020bze, Barman:2021lot, Nath:2021uqb, Das:2021nqj, Barman:2021yaz}. In this context we recently proposed a prescription, dubbed as the\textit{Scale Invariant FIMP Miracle}, where we showed in a simple scale-invariant Higgs-portal vector DM framework, due to constrained relations among the quartic and gauge couplings the mixing in the scalar sector is always set to be a fixed number $\sin\theta\sim\mathcal{O}(10^{-5})$ to produce the observed DM abundance via freeze-in~\cite{Barman:2021lot}. Moreover, in such a framework, direct detection of freeze-in DM~\cite{Hambye:2018qjv, Heeba:2019jho, Elor:2021swj} also becomes important, again due to the underlying scale invariance of the theory that provides a naturally light non-standard Higgs that can also be searched for in several lifetime/intensity frontier facilities. 

In this paper we explore a cosmological option for freeze-in DM candidates that can give rise to observed relic abundance for large couplings with the visible sector, and thereby boosting the detection prospects\footnote{DM genesis in a modified cosmological scenario, under the influence of an early matter era has been discussed in~\cite{Bernal:2018ins, Bernal:2018kcw, Cosme:2020mck}.}. Traditionally, DM production in the early Universe relies on the central assumption that at the time of DM production, the energy budget of the Universe was dominated by radiation. We know from Big Bang Nucleosynthesis (BBN) that this is definitely the case at temperatures around and below $T_{\rm BBN} \simeq \text{few} \; {\rm MeV}$~\cite{Kawasaki:2000en,Ichikawa:2005vw}. However, we have no direct information about the energy budget of the Universe at very early times i.e., pre-BBN epoch. The WIMP or FIMP relic may differ by orders of magnitude if deviations from a standard cosmological history are considered.  Following the prescription in~\cite{DEramo:2017ecx, DEramo:2019tit}, in order to investigate freeze-in DM production in non-standard cosmological era,  a species $\varphi$ is introduced with the following characteristic of energy density red-shifting with the scale factor $a$ as
\be
\rho_\phi \propto a^{-(4+n)} \ \,(n > 0)\,,
\label{eq:phiScaling}
\ee
where $n=0$ corresponds to the case of standard radiation-dominated era. We also define $T_R \gtrsim T_{\rm BBN}$ as the temperature of transition from early matter domination era (due to $\varphi$) to radiation-domination era. Hence, in the independent two-dimensional parameter space $\{T_R, n\}$ of all possible cosmological backgrounds, we investigate the DM production. As expected, due to the presence of a new species, the Hubble parameter is always larger than that in the radiation-dominated-era and the Universe expands {\em faster} when at temperatures $T>T_R$. As it has already been established in~\cite{ DEramo:2019tit}, in a faster-than-usual expanding Universe, freeze-in production is suppressed, implying that to produce enough DM to match the observations, larger couplings, and thus larger detection rates, are in order\footnote{For freeze-out phenomenology in a fast expanding Universe see, for example, Refs.~\cite{Barman:2021ifu, Arcadi:2021doo}.}. In the same spirit, for the scale-invariant vector DM model discussed in Ref.~\cite{Barman:2021lot}, we show the scalar mixing $\sin\theta$ (or equivalently the DM-SM coupling) can be improved by several orders of magnitude with respect to the standard radiation domination once a faster expansion is invoked. This, as one can understand, drastically improves the detection prospects of the freeze-in DM in different experimental facilities. We also show very large $n$ values which require larger DM-SM coupling to produce the right abundance, are not only discarded from the condition of non-thermal DM production (depending on the choice of $T_R$), but also from existing experimental bounds. We thus turn the argument around and constrain physically realizable cosmological models which can give rise to faster expansion rate by remaining well within experimental limits.

The paper is organized as follows: in Sec.~\ref{sec:model} we briefly describe the scale-invariant Higgs-portal vector dark matter model, we then motivate a faster cosmological expansion history of the Universe in Sec.~\ref{sec:freeze-in-fast-exp}, in Sec.~\ref{sec:lab-probe} we investigate various collider and/or low energy experimental probes of the model and finally we conclude in Sec.~\ref{sec:concl}.

\section{Model Set-up}
\label{sec:model}
Here we describe the scale invariant Higgs-portal abelian vector dark matter model briefly (the model has been introduced in detail in~\cite{Barman:2021lot}). The SM gauge symmetry is augmented with an abelian $\ux$ gauge symmetry and a complex scalar $S$, such that under a discrete $\mathbb{Z}_2$ symmetry the new fields transform as
\begin{equation}
X_\mu\to-X_\mu;~~S\to S^\star\,,           
\end{equation}
while all the SM fields are even under the $\mathbb{Z}_2$. The interaction Lagrangian therefore turns out to be
\begin{equation}
\mathcal{L}\supset -\frac{1}{4}X_{\mu\nu}X^{\mu\nu}+\left|D_\mu S\right|^2-V\left(H,S\right)\,, 
\end{equation}
where $H$ is the SM Higgs doublet 
and $D_\mu = \partial_\mu + ig_X\,X_\mu$  is the covariant derivative. Next, we consider the tree-level renormalizable scale-invariant scalar potential as
\begin{equation}
V\left(H,S\right)_\text{cl}=\lambda_H\,|H|^4+\lambda_S\,\left|S\right|^4-\lhs\,|H|^2\,\left|S\right|^2\,, 
\label{eq:tree-pot}
\end{equation}
which shows that the dark and the visible sectors communicate via the renormalizable $\lhs$ coupling with the limit $\lhs\to 0$  corresponds to the case where the hidden sector completely decoupled from the visible sector.  
As the classical potential is zero along the so-called ``flat direction", the one-loop correction necessarily dominates there. Since at the minimum of the 1-loop effective potential $V_\text{cl}\geq 0$ and $V_\text{eff}^\text{1-loop}<0$, the minimum of $V_\text{eff}^\text{1-loop}$ along the flat direction, where $V_\text{cl}=0$, is a global minimum of the full potential~\cite{YaserAyazi:2019caf,Mohamadnejad:2019vzg}. Therefore, spontaneous symmetry breaking (SSB) indeed occurs and we expand the fields around the minima 
\begin{equation}
S=\frac{1}{\sqrt{2}}\left(s+v_s\right)\,,H=\frac{1}{\sqrt{2}}\Big\{0,h+v_h\Big\}^T\,, \label{eq:vev-exp}
\end{equation}
where the two VEVs are related via
\begin{equation}
\frac{v_h^2}{v_s^2}=\frac{\lhs}{2\lambda_H}\,,
\label{eq:vev}    
\end{equation}
defining the ``flat direction", along which $V_\text{cl}=0$. As $S$ receives a non-vanishing VEV, the mass of the $\ux$ gauge boson i.e., the DM can be expressed as 
\begin{equation}\begin{aligned}
& m_X =g_X v_s\equiv g_X\,\sqrt{2\lambda_H/\lhs}\,v_h\,.
\end{aligned}\label{eq:mx}
\end{equation}
To calculate the tree-level masses, we construct the mass matrix in the weak basis as
\begin{equation}
\mathcal{M}^2=\left(\begin{array}{cc}
 2 v_h^2 \lambda_H & -v_h v_s \lambda_{HS} \\
 -v_h v_s \lambda_{HS} & 2 v_s^2 \lambda_s\\
\end{array}\right).
\end{equation}
We can then rotate it to the physical (mass) the basis via
\begin{equation}
\begin{pmatrix}
h_1 \\ h_2 
\end{pmatrix} = \begin{pmatrix}
\cos\theta & -\sin\theta\\\sin\theta & \cos\theta
\end{pmatrix}\,\begin{pmatrix}
h \\ s
\end{pmatrix}
\end{equation}
where the mixing angle is given by
\begin{equation}
\tan2\theta = \frac{2v_h\,v_s}{v_h^2-v_s^2}\implies\sin\theta=\frac{v_h}{\sqrt{v_h^2+v_s^2}}\approx\frac{v_h}{v_s} 
\label{eq:mixing}
\end{equation}
for $v_s\gg v_h$. In tree-level, the field $h_2$ being along the flat direction, is massless\footnote{This field with vanishing zeroth-order mass, is dubbed as ``scalon"~\cite{Gildener:1976ih}.}. This acquires a radiative mass along the flat direction à la Coleman-Weinberg~\cite{PhysRevD.7.1888}, and  becomes a pesudo-Nambu-Goldstone boson (pNGB) at quantum level. The field $h_1$, on the other hand, is perpendicular to the flat direction which we identify as the SM-like Higgs, observed at the LHC with a mass of $m_{h_1}=125$ GeV~\cite{Mohamadnejad:2019vzg}.

As mentioned above, the scale invariance of the theory gives rise to massless scalar field in the classical level. One loop correction then breaks the scale invariance giving mass to the massless eigenstate $h_2$. Following~\cite{Gildener:1976ih}, the 1-loop effective potential can be written as~\cite{Foot:2007as, Foot:2007iy,Farzinnia:2013pga,YaserAyazi:2019caf,Mohamadnejad:2019vzg, Kannike:2020ppf}
\begin{equation}
V_\text{eff}^\text{1-loop} = \alpha h_2^4+\beta h_2^4\log\frac{h_2^2}{\mu^2}\,,
\label{eq:1-loop}
\end{equation}
with $\alpha,\beta$ as dimensionless constants defined as
\begin{equation}
\alpha=\frac{1}{64\pi^2\,v^4}\sum_j g_j m_j^4\log\frac{m_j^2}{v^2}\,,\beta=\frac{1}{64\pi^2\,v^4}\sum_j\,g_j\,m_j^4\,,
\end{equation}
where $g_j$ and $m_j$ are the tree-level mass and the internal degrees of freedom of the $j^\text{th}$ particle, $v^2=v_s^2+v_h^2$ and $\mu$ is the renormalization scale. Minimizing Eq.~\eqref{eq:1-loop} we find that the potential has a non-trivial stationary point at~\cite{Foot:2007as,Farzinnia:2013pga}
\begin{equation}
\mu = v\,\exp\Biggl(\frac{\alpha}{2\beta}+\frac{1}{4}\Biggr).
\label{eq:1-loop-min}
\end{equation}
The 1-loop potential can now be re-written utilizing Eq.~\eqref{eq:1-loop-min} as
\begin{equation}
V_\text{eff}^\text{1-loop} = \beta\,h_2^4\Biggl[\log\frac{h_2^2}{v^2}-\frac{1}{2}\Biggr]\,.
\end{equation}
With this we can now express the mass of $h_2$ as~\cite{PhysRevD.7.1888,Gildener:1976ih}
\begin{equation}
m_{h_2}^2 = \frac{d^2V_\text{eff}^\text{1-loop}}{d h_2^2}\Big|_v=8\,\beta\, v^2\,,    
\end{equation}
where again we see for $\beta>0$ this is positive definite. Taking all standard and non-standard particles into account, we obtain
\begin{equation}
m_{h_2}^2  = \frac{v_h^4}{8\pi^2 v^2}\Bigl(\lh^4+\frac{3}{8}g_{2}^4+\frac{3}{16}g_{2}^4\left(g_2^2+g_1^2\right)^2+3 g_X^4\left(v_s/v_h\right)^4-3y_t^4\Bigr)
\label{eq:mh2}
\end{equation}
where $g_2$ and $g_1$ are the gauge coupling corresponding to the SM groups $SU(2)_L$ and $U(1)_Y$ gauge respectively, and $y_t$ is the SM top Yukawa coupling. Also note, the fermion contribution appears with a relative sign. From Eq.~\eqref{eq:mh2} it is very important to note the role of the $U(1)_X$ gauge boson, without which $m_{h_2}^2<0$ (or equivalently $\beta<0$). This also puts a lower bound on the DM mass in the present model, requiring $m_X\gtrsim 240$ GeV.

\section{Freeze-in in the era of Fast Expansion}
\label{sec:freeze-in-fast-exp}
\subsection{Fast expansion: summary}
\label{sec:fast-exp}
We assume the Universe before BBN has two different species: radiation and some other species $\varphi$ with  energy densities $\rho_{\rm rad}$ and $\rho_\varphi$ respectively. In presence of a new species ($\varphi$) along with the radiation field, the total energy budget of the Universe is $\rho = \rho_{\text{rad}} + \rho_\varphi$. One may always express the energy density of the radiation component as function of temperature $T$ as $\rho_{\text{rad}} (T) = \frac{\pi^2}{30}\,g_{\star}(T)T^4$, with $g_{\star}(T)$ being the effective number of relativistic degrees of freedom at temperature $T$. In the absence of entropy production per comoving volume i.e., $sa^3=$ const., radiation energy density redshifts as $\rho_{\text{rad}}(t)\propto a(t)^{-4} $. In case of a rapid expansion of the Universe the energy density of $\varphi$ field is expected to be redshifted faster than the radiation. Accordingly, one can assume $\rho_\varphi\propto a(t)^{-(4+n)}$, where $n>0$ implies $\varphi$ energy density dominates over radiation during early enough times. Now, the entropy density of the Universe is expressed as $s(T) = \frac{2\pi^2}{45}g_{*s}(T)T^3$, where $g_{*s}\left(T\right)$ is the effective relativistic degrees of freedom. A general form of $\rho_\varphi$ can then be constructed using the entropy conservation $g_\star\left(T\right)^{1/3}aT=\text{constant}$ in a comoving frame as
\begin{equation}
 \rho_\varphi(T) =  \rho_\varphi(T_R)\left(\frac{g_{*s}(T)}{g_{*s}(T_R)}\right)^{(4+n)/3}\left(\frac{T}{T_R}\right)^{(4+n)}\,,
\end{equation}
where the temperature $T_R$ is an unknown variable and can be considered as the point of equality where $\rho_\varphi(T_R)=\rho_{\text{rad}}(T_R)$ is achieved. Using this, the total energy density at any temperature $T$ can be expressed as
\begin{align}\label{eq:totalrho}
 \rho(T) &= \rho_{rad}(T)+\rho_{\varphi}(T)=\rho_{rad}(T)\left[1+\frac{g_* (T_R)}{g_* (T)}\left(\frac{g_{*s}(T)}{g_{*s}(T_R)}\right)^{(4+n)/3}\left(\frac{T}{T_R}\right)^n\right]\,.
\end{align}
Now, from standard Friedman equation one can write the Hubble parameter in terms of the energy density as
\begin{equation}  
\mathcal{H} = \frac{\sqrt{\rho}}{\sqrt{3}M_P}\,,
\end{equation}
with $M_P$ being the reduced Planck mass. At temperature higher than $T_R$ with $g_*(T) = \bar g_*$ (some constant), the Hubble rate can approximately be recasted as
\begin{equation}\begin{aligned}
& \mathcal{H}(T)= \mathcal{H}_R\left(T\right)\Biggl[1+\Biggl(\frac{T}{T_R}\Biggr)^n\Biggr]^{1/2} \approx \frac{\pi\bar g_*^{1/2}}{3\sqrt{10}}\, \frac{T^2}{M_P}\,\left(\frac{T}{T_R}\right)^{n/2}\,(T \gg T_R)\,,
 \label{eq:mod-hubl}      
    \end{aligned}
\end{equation}
where $H_R\left(T\right)$ is the Hubble parameter in the standard radiation dominated Universe. In case of the SM, $\bar g_*\equiv g_*\text{(SM)} = 106.75$. It is important to note from Eq.~\eqref{eq:mod-hubl} that the expansion rate is larger than what it is supposed to be in the standard cosmological background for $T>T_R$ and $n>0$. The temperature $T_R$ can not be too small such that it changes the standard BBN history, rather it follows
\begin{equation}
T_R \gtrsim \left(15.4\right)^{1/n}~\text{MeV}\,,
\label{eq:tr-bbn}
\end{equation}
as derived in Appendix.~\ref{sec:bbn}. As one can see, for larger $n$ this constraint becomes more and more lose. It is important to note here that the above bound seems ill-defined for $n=0$, simply because for $n=0$ there is no bound from $\Delta N_\text{eff}$ at all due to the absence of the extra degrees of freedom $\varphi$ around BBN, hence this bound ceases to exist for $n=0$.

Finally, we would like to briefly touch upon some of the physical realizations of faster than usual expansion history of the Universe. So far in the analysis we have assumed $T_R\,,n$ to be free parameters without going into the details of the properties of the new species $\varphi$. Assuming $\varphi$ to be a real scalar field minimally coupled to gravity~\cite{DEramo:2017ecx, DEramo:2019tit}, we find that the energy density of such a species redshifts as
\begin{equation}
\rho_\varphi\propto a^{-3\left(1+\omega\right)}\,,
\end{equation}
where we get
\begin{equation}
\omega = \frac{1}{3}\left(n+1\right)\,,
\end{equation}
on comparing with $\rho_\varphi\propto a^{-\left(4+n\right)}$. For a positive scalar potential, $\omega\in\left[-1,+1\right]$ that leads to $n\in\left[-4,2\right]$. Theories with $n=2$ are quintessence fluids motivated by the accelerated expansion of the Universe~\cite{Caldwell:1997ii, Sahni:1999gb}. One possible realization of scalar potential giving rise to this behaviour is $V(\varphi)\sim \exp\left(-\lambda\,\varphi\right)$~\cite{PhysRevD.37.3406, Copeland:1997et, Muromachi:2015nva}. For theories with $n>2$ one has to consider scenarios faster than quintessence. Example of such theories can be found, for example, in~\cite{Buchbinder:2007ad}, where one assumes the presence of a pre big bang ``ekpyrotic" phase. The key ingredient of ekpyrosis is same as that of inflation, namely a scalar field rolling down some self-interaction potential. However, the crucial difference being, while inflation requires a flat and positive potential, its ekpyrotic counterpart is steep and negative. Other than this version of the non-standard cosmology, there is also the scope for modification of Hubble rate at sub-horizon scales, due to modified gravity theories particularly considering Modified gravity dominated eras during the pre-BBN epoch~\cite{Dunsby:1998hd, Lin:2016gve, Nunes:2018zot,Modak:1999nm, Schelke:2006eg, Catena:2009tm, Dent:2009bv, Leon:2013qh}. Formation of cosmological relics in modified gravities with modified Hubble expansion rate were considered in~\cite{Modak:1999nm, Schelke:2006eg, DAmico:2009tep, Catena:2009tm, Dent:2009bv, Leon:2013qh}. Particularly, motivated from Randall-Sundrum type II brane cosmology~\cite{Randall:1999vf},  kination models~\cite{Salati:2002md, Pallis:2005hm, Guo:2009nt}, from dark sector additional degrees of freedom in the thermal plasma leading to boost factor~\cite{Catena:2009tm}, and from
 cosmology with $f(x)=x+\alpha\,x^n$, where $x=R$, ${\cal T}$; $R$ and ${\cal T}$ being the scalar curvature and the scalar torsion, respectively~\cite{Capozziello:2008rq, Capozziello:2015ama, Cai:2015emx, Capozziello:2017bxm} all of which explicitly include non-standard cosmological era prior to BBN. In our analysis although we do not consider the questions related to primordial cosmology like inflation or bounce or any spectator however in principle the origin of the scalar field is associated to such primordial era as well, either in the form of a modification of the geometry or as a modification to the matter sector. Our analysis and results are presented in terms of the general parametrization we mentioned and the results we get will be applicable to all such scenarios in general.
 
 %

\subsection{Modified freeze-in yield}
\label{sec:dm-yield}
\begin{figure}[htb!]
    \centering
    \includegraphics[scale=0.35]{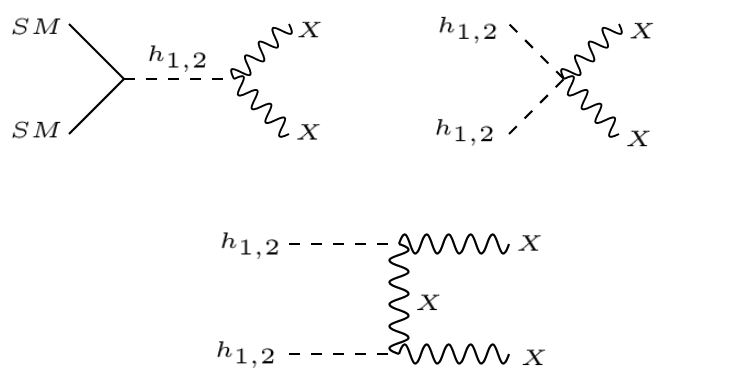}
    \caption{ \it Relevant Feynman diagrams for DM production via freeze-in.}
    \label{fig:feyn}
\end{figure}

The key for freeze-in is to assume the DM was absent in the early Universe and then the DM abundance gradually builds up from the thermal bath with time. Since the scale invariance of the model necessarily fixes the DM mass $m_X\gtrsim 240$ GeV, hence the only possible way to produce DM in the eraly Universe via freeze-in is through 2-to-2 scattering of the bath particles, mediated via (non-) standard Higgs. All the relevant Feynman graphs for DM production are shown in Fig.~\ref{fig:feyn}\footnote{Since $m_{h_2}\ll m_X$, hence $h_2$ influences the resulting DM abundance very weakly~\cite{Duch:2017khv}.}. The evolution of the DM number density with time can be tracked by solving the Boltzmann equation (BEQ), which in the present scenario reads
\begin{equation}\label{eq:beq}
x\,\mathcal{H}\,s\,\frac{dY_X}{dx} = \gamma_\text{ann}\,,
\end{equation}
where
\begin{equation}\begin{aligned}
\gamma\left(a,b\to1,2\right)&=\int\prod_{i=1}^4 d\Pi_i \left(2\pi\right)^4 \delta^{(4)}\biggl(p_a+p_b-p_1-p_2\biggr)f_a{^\text{eq}}f_b{^\text{eq}}\left|\mathcal{M}_{a,b\to1,2}\right|^2\\&=\frac{T}{32\pi^4}g_a g_b \int_{s_{min}}^\infty ds~\frac{\biggl[\bigl(s-m_a^2-m_b^2\bigr)^2-4m_a^2 m_b^2\biggr]}{\sqrt{s}}\sigma\left(s\right)_{a,b\to1,2}K_1\left(\frac{\sqrt{s}}{T}\right)\label{eq:gam-ann}\,, \end{aligned}    \end{equation}
with $a,b(1,2)$ as the incoming (outgoing) states and $g_{a,b}$ are corresponding degrees of freedom. Here $f_i{^\text{eq}}\approx\exp^{-E_i/T}$ is the Maxwell-Boltzmann distribution. The Lorentz invariant 2-body phase space is denoted by: $d\Pi_i=\frac{d^3p_i}{\left(2\pi\right)^3 2E_i}$. The amplitude squared (summed over final and averaged over initial states) is denoted by $\left|\mathcal{M}_{a,b\to1,2}\right|^2$ for a particular 2-to-2 scattering process. The lower limit of the integration over $s$ is $s_{min}=\text{max}\biggl[\left(m_a+m_b\right)^2,\left(m_1+m_2\right)^2\biggr]$. Here we define the DM yield $Y_X=n_X/s$ as the ratio of DM number density to the comoving entropy density in the visible sector since the DM is only produced from the SM bath. The parameter $x=m_X/T$ describes the SM sector temperature $T$, the modified Hubble parameter is defined as in Eq.~\ref{eq:mod-hubl}, and $\gamma=\langle\sigma v\rangle n_\text{eq}^2$ is the reaction density~\cite{Chu:2011be} for the SM particles annihilating into the DM. We have collected all the 2-to-2 cross-sections as a function of the center of mass energy $\sqrt{s}$ in Appendix~\ref{sec:app-ann}\footnote{In general, any particle that couples in the thermal bath with the primordial plasma is expected to obtain a mass proportional to the temperature of the Universe provided the condition $T>m$ is satisfied~\cite{Laine:2016hma, Konar:2021oye, Chakrabarty:2022bcn}. However, here we are interested in IR freeze-in, where the DM yield becomes important at lower temperature (cf. Fig.~\ref{fig:dm-yld}), thus the effect of such corrections are rather small, and hence the rates given in Appendix~\ref{sec:app-ann} are still valid.}. The relic abundance of the DM at the present epoch is computed using
\begin{equation}
\Omega_X h^2 = \left(2.75\times 10^8\right) \left(\frac{m_X}{\text{GeV}}\right) Y_X(T_0)\,,
\label{eq:relicX}    
\end{equation}
where $T_0$ is the present temperature of the Universe. 

\begin{figure}[htb!]
    \centering
    \includegraphics[scale=0.37]{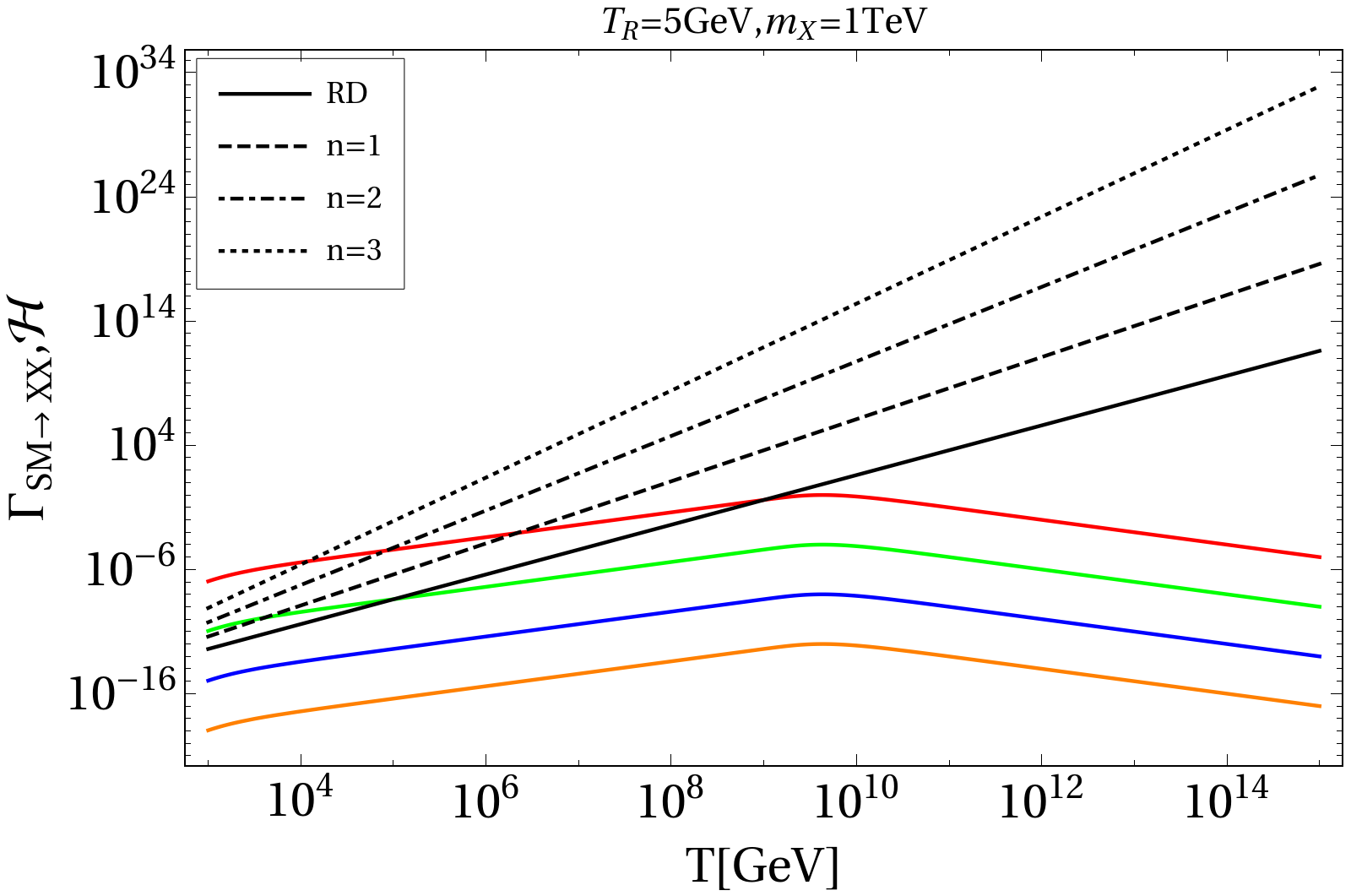}~~~~\includegraphics[scale=0.37]{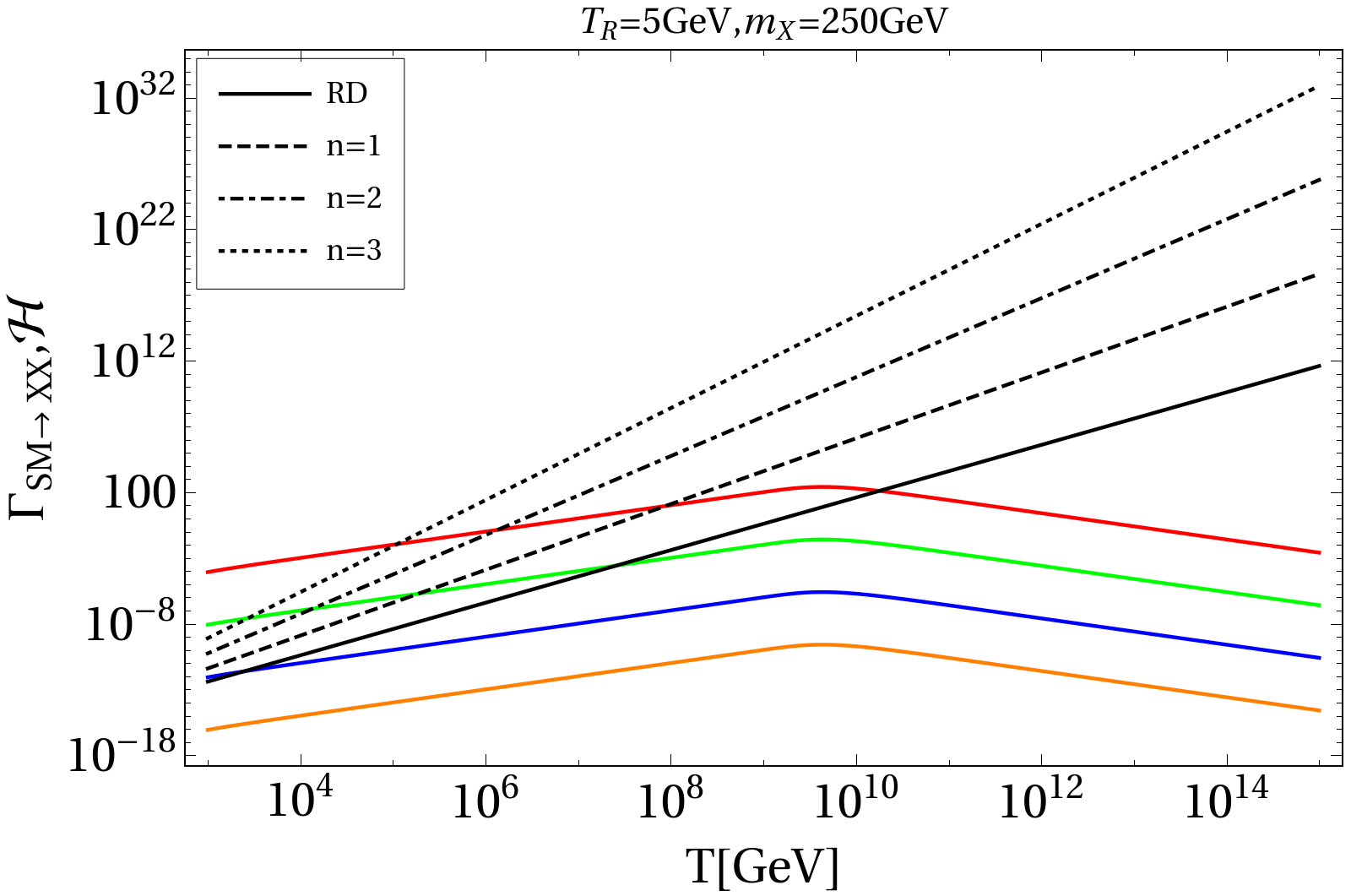}
    \caption{ \it Rate of DM interaction as a function of bath temperature $T$ for DM of mass 1 TeV (left panel) and 250 GeV (right panel) with a fixed $T_R=5$ GeV. Different coloured curves correspond to $g_X=\{10^{-4},\,10^{-3},\,10^{-2},\,10^{-1}\}$ from bottom to top.}
    \label{fig:rate}
\end{figure}

\begin{figure}[htb!]
    \centering
    \includegraphics[scale=0.37]{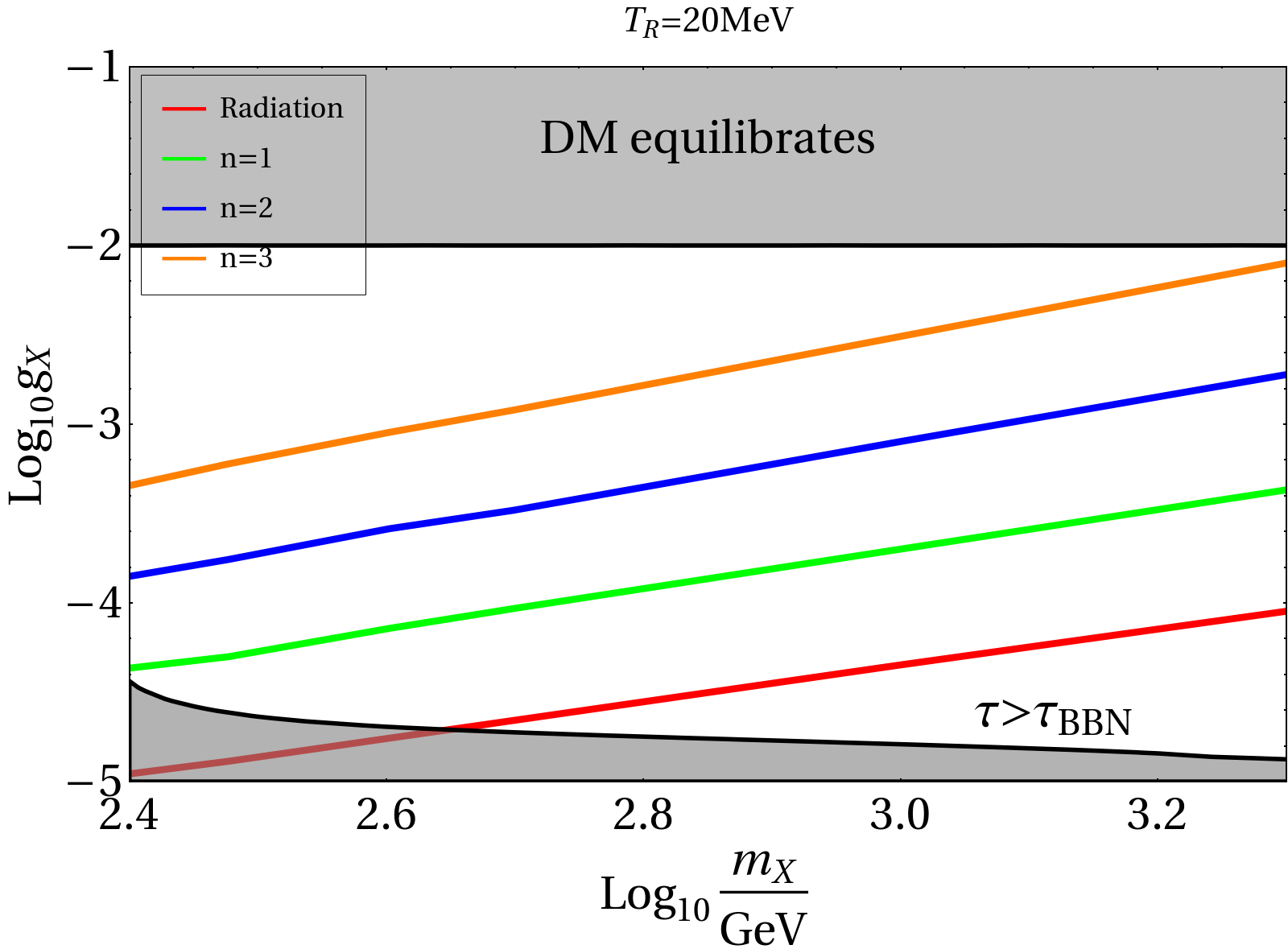}~~~~\includegraphics[scale=0.37]{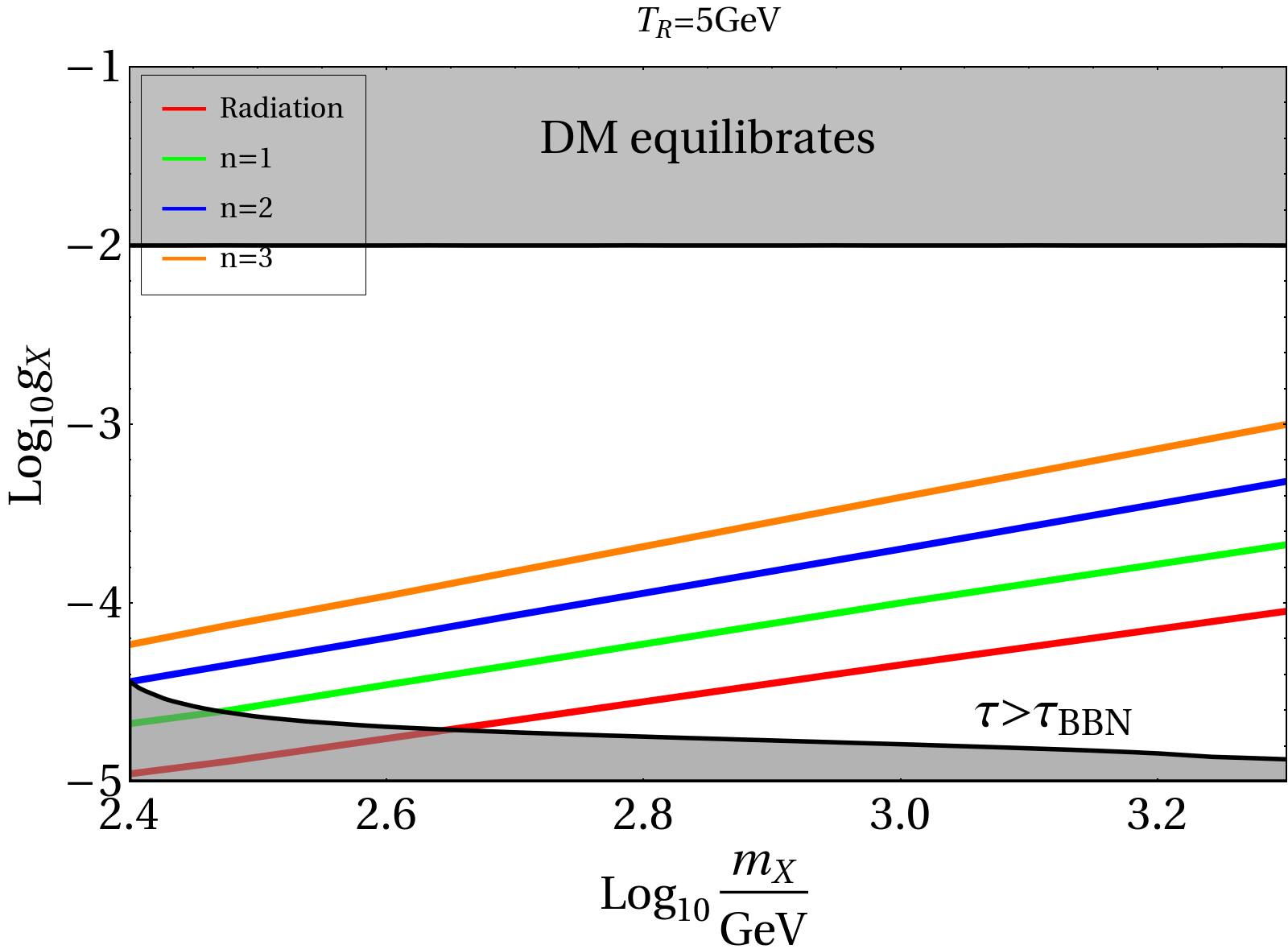}
    \caption{ \it PLANCK observed relic density allowed parameter space in the bi-dimensional plane of $g_X-m_X$, where in the left panel we have $T_R=20$ MeV and in the right panel $T_R=5$ GeV. Different coloured lines correspond to different choices of $n=\{0,\,1,\,2,\,3\}$ in red, green, blue and orange respectively. The gray shaded regions are excluded from BBN bound on the light scalar lifetime (bottom) and from the requirement of non-thermal DM production (top). }
    \label{fig:gx-mx}
\end{figure}

The evolution of DM yield with temperature are shown in Fig.~\ref{fig:dm-yld} for a fixed DM of mass 250 GeV. In the left panel we show the behaviour of the yield for different choices of $n$, where $n=0$ indicates simple radiation dominated background (red curve). We have also fixed $T_R=10$ MeV, which is above the BBN temperature $(\sim 1\,\text{MeV})$. Here all the curves satisfy the central value of the PLANCK observed relic density. We see, with a larger $n$, in order to satisfy the relic abundance, a larger coupling is needed. This is a general trend, i.e., in a faster than usual expanding Universe the DM is always under produced compared to the case of a standard history. This pattern can be explained by considering an approximate analytical solution for the asymptotic DM yield
\begin{align}
& Y_X(\infty)\approx\frac{135\,\sqrt{10}}{g_{\star s}\,\sqrt{g_{\star\rho}}\,\pi^7}\,\frac{g_X^4\,M_P}{m_X}\,\left(\frac{2}{x_R}\right)^\frac{n}{2}\,\Gamma\left[\frac{n+2}{4}\right]\,\Gamma\left[\frac{n+10}{4}\right]\,,      
\end{align}
where we have introduced $x_R=m_X/T_R$ and $\Gamma(x)$ is the usual gamma function. We have also assumed the 2-to-2 cross-section goes as $\sigma(s)\sim g_X^4/s$ (dimensionally), which leads to the reaction density $\gamma_\text{ann}\simeq\frac{g_X^4}{4\,\pi^4}\,m_X^2\,T^2\,K_2\left(2\,x\right)$, with $x=m_X/T$. Note that, the DM yield is modified by a suppression factor
\begin{align}
& Y_X(\infty) \simeq  Y_X^\text{rad}(\infty)\times\Biggl(\frac{4}{3\,\pi}\,\left(\frac{2}{x_R}\right)^{\frac{n}{2}}\,\,\Gamma\left[\frac{n+2}{4}\right]\,\Gamma\left[\frac{n+10}{4}\right]\Biggr)\,,
\end{align}
\begin{figure}[htb!]
    \centering
    \includegraphics[scale=0.37]{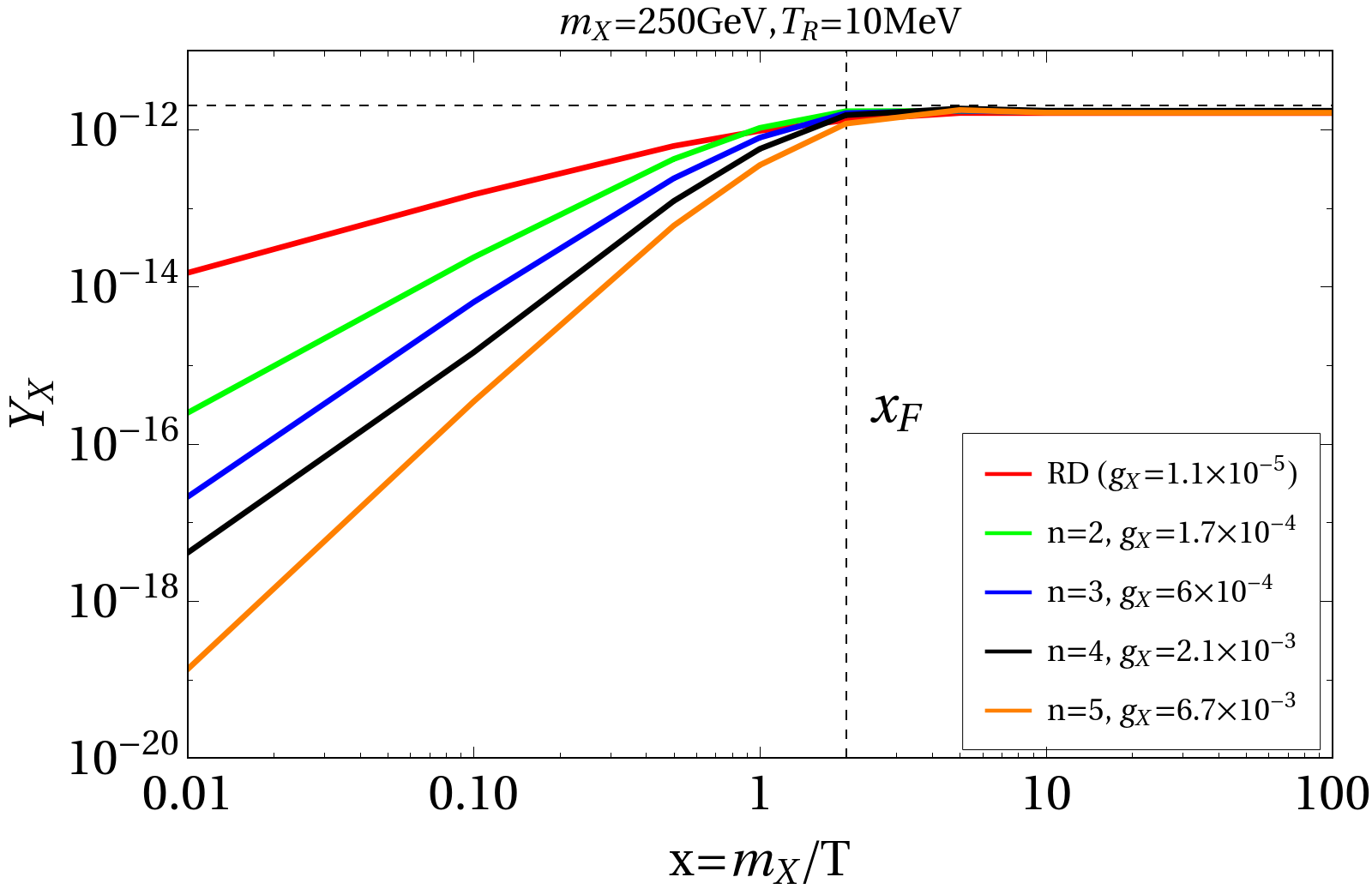}~~~~ \includegraphics[scale=0.37]{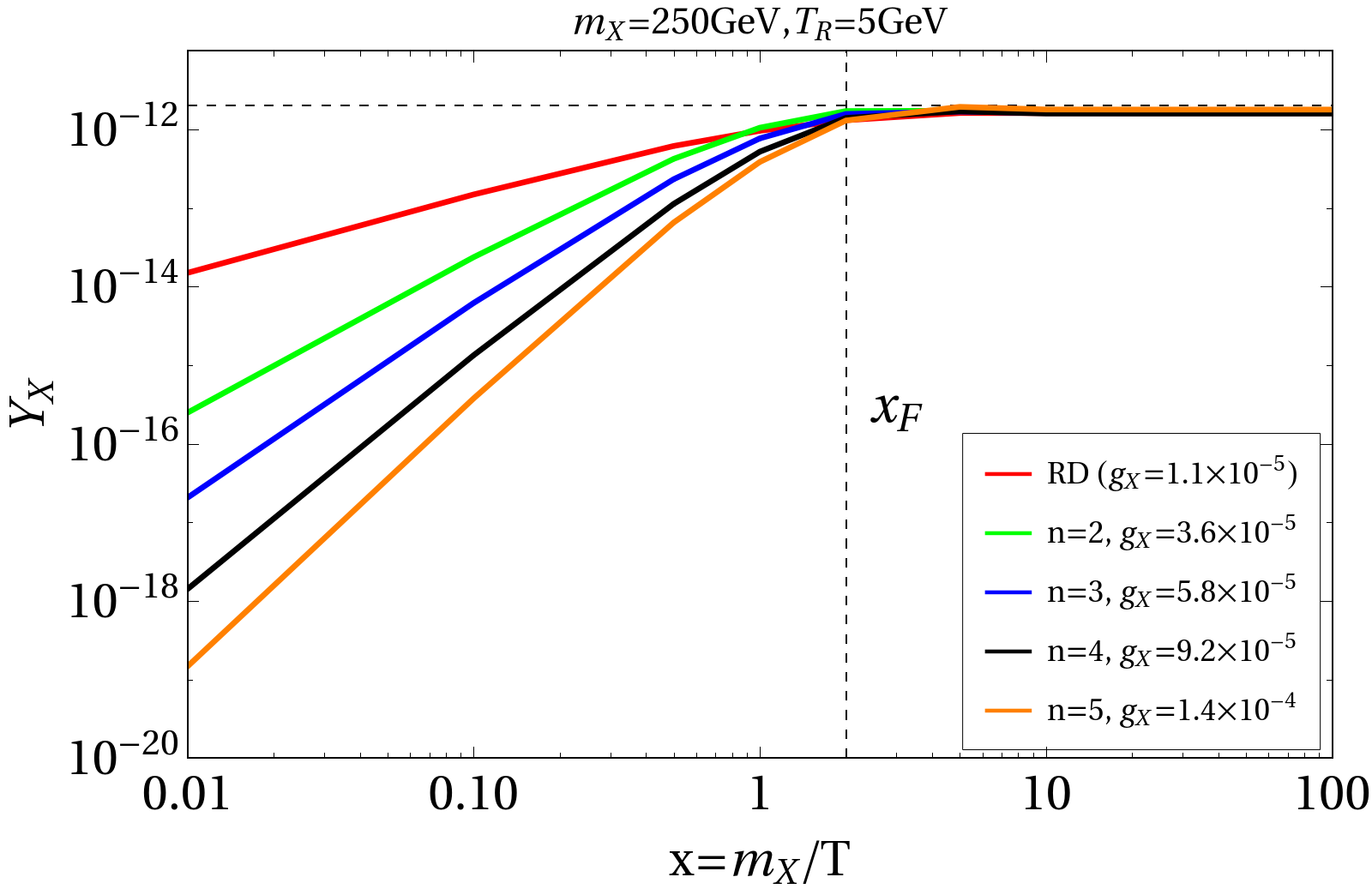}
    \caption{ \it Evolution of DM yield with $x=m_X/T$, where all curves satisfy relic abundance for different choices of the gauge coupling $g_X$, for different expansion parameter $n$ with fixed DM mass $m_X=250$ GeV and a fixed $T_R=$ 10 MeV(left) and $T_R=$ 5 GeV (right). Here $x_F$ denotes the temperature at which DM yield saturates.}
    \label{fig:dm-yld}
\end{figure}
which implies, for any $n>0$, in order to produce the observed relic abundance, one has to choose a larger $g_X$ compared to the standard radiation dominated Universe, for a given DM mass. This is exactly what is reflected in the left panel of Fig.~\ref{fig:dm-yld}, where we see approximately $\sim\mathcal{O}(100)$ improvement in the new gauge coupling compared to the standard cosmological history (in red) when we choose a faster expansion with $n=5$ (in orange). In the right panel we show the same, but for a larger $T_R=5$ GeV. Here we see the improvement in $g_X$ compared to the radiation domination is $\sim\mathcal{O}(10)$, as larger $T_R$ tends towards standard radiation domination. Thus, a larger $n$ and smaller $T_R$ is preferred to see the effect of modified cosmological history in the present set-up. However, one has to keep in mind that it is not possible to increase $g_X$ incessantly since larger $g_X$ shall end up thermalizing the DM, which is not desirable in the freeze-in paradigm. Since in the present case the DM interaction rate scales as $g_X^4$, hence it is rather easier for the DM to remain out of equilibrium for the size of $g_X$ required to produce right abundance via freeze-in. To ensure the DM production is non-thermal in the early Universe, we compare the DM-SM interaction rate $\Gamma_\text{int}=\left(n_\text{eq}\right)_j\,\langle\sigma v\rangle^j$ ($j\in$ SM) with the modified Hubble parameter (cf.Eq.~\eqref{eq:mod-hubl}). For a DM of mass 1 TeV (in the left panel of Fig.~\ref{fig:rate}), in the standard cosmological background, to ensure non-thermal DM production, one has to choose $g_X\lesssim 10^{-2}$ such that the DM remains out of equilibrium till $T\sim 1$ TeV. In a modified cosmological scenario, with $n\lesssim3$, the bound on $g_X$ can be relaxed and $g_X\lesssim 10^{-1}$ can still ensure the interaction rate to remain below the Hubble rate till $T\sim 1$ TeV. For $n>3$, the bound on $g_X$ can be further relaxed as shown by the black tilted lines in Fig.~\ref{fig:rate}. This is expected, since a larger $n$ essentially leads to a faster Hubble rate making it easier for the DM to stay out of equilibrium as the temperature drops. Note that, for heavier DM it is easier to satisfy the out of equilibrium condition as the interaction cross-section becomes smaller due to phase space suppression. On the other hand, for the minimum allowed DM mass (due to scale invariance) $m_X\simeq 250$ GeV, as one can see from the right panel of Fig.~\ref{fig:rate}, $g_X=10^{-2}$ makes the DM thermal at $T\sim 10^6$ GeV for standard radiation domination, which drastically improves with $n=1$. Thus, irrespective of the DM mass, we put a conservative bound on the gauge coupling $g_X<10^{-2}$, since we confine ourselves within $n\leq 4$ for the phenomenological study. 
The relic density allowed parameter space can be portrayed in the $g_X-m_X$ plane, as shown in Fig.~\ref{fig:gx-mx}. Here we see the typical feature of scale invariance, i.e., with increase in DM mass one needs a larger $g_X$ to satisfy the relic constraint $\Omega_X h^2\simeq 0.12$. This typical nature is attributed to the fact that the relic abundance goes as $\Omega_X h^2\propto m_X\times Y_X \propto \left(g_X/m_X\right)^4$, where $Y_X \sim \sigma\left(T_\text{FI}\right)\, M_P\,T_\text{FI}\sim \frac{g_X^4 M_P}{m_X^5}\,m_{h_1}^4$ is the approximate IR yield of the DM~\cite{Duch:2017khv, Barman:2021lot}, with $T_\text{FI}\sim m_X$ being the temperature at which the DM abundance saturates. Again, a larger $n$ requires a larger $g_X$ to obtain the right abundance as we have already seen. The presence of the decaying light scalar $h_2$ poses a very important bound on the DM parameter space. Since $h_2$ decay to SM particles (see Appendix.~\ref{sec:app-h2-decay}) is suppressed by the small mixing angle, its lifetime tends to be very long. Particularly, there will be an upper bound on the $h_2$-lifetime from nucleosynthesis. A fully quantitative analysis of these effects is beyond the scope of this paper. Instead we follow~\cite{Kawasaki:2000en} and, to remain within the 2$\sigma$ limit of the observed $\textsuperscript{4}\text{He}$ abundance, require $\Gamma_{h_2}^{-1}\equiv\tau_{h_2}<1$ sec. This is rather a conservative bound given the fact that the light scalar never comes in equilibrium with the SM due to the feeble portal coupling $\lhs$. Note that, a larger $n$ can save the parameter space from the BBN bound, but can lead to thermalization of the DM in the early Universe, depending on the choice of $T_R$. 

\begin{figure}[htb!]
    \centering
    \includegraphics[scale=0.4]{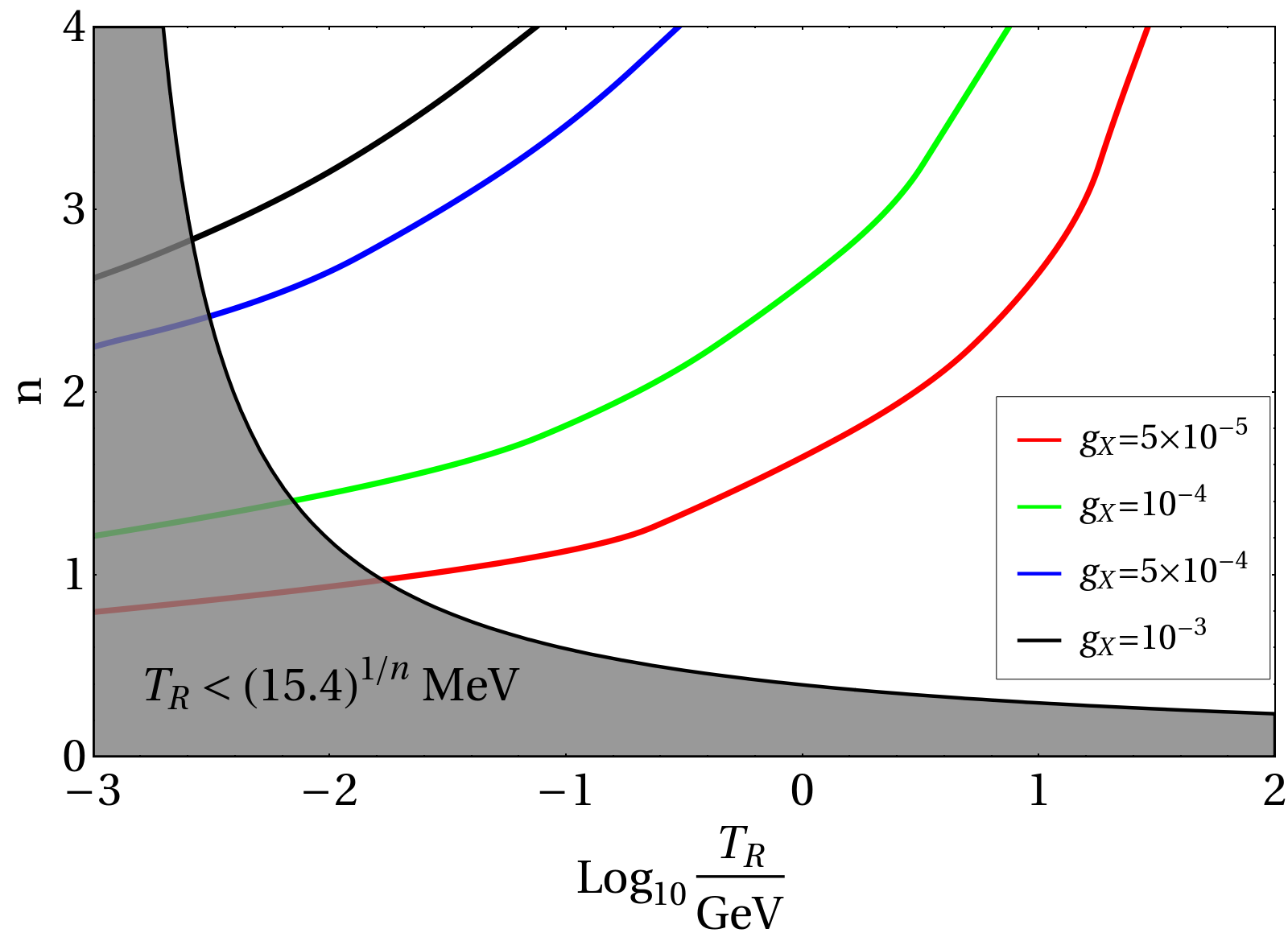}
    \caption{\it Contours with different colours  satisfy the PLANCK observed relic abundance for different choices of $g_X$ (as mentioned in the plot label), for a DM of mass 300 GeV. The shaded region is disallowed from the BBN constraint on $T_R$ (cf. Eq.~\eqref{eq:tr-bbn}). }
    \label{fig:ntr}
\end{figure}
We finally show the relic density allowed parameter space in $n-T_R$ plane for a fixed DM mass of 300 GeV. A larger $T_R$ results in a reduced Hubble rate (cf.Eq.~\eqref{eq:mod-hubl}), that has to be compensated with a larger $n$ to produce the observed relic abundance. However, The constraint arising from the BBN bound due to Eq.~\eqref{eq:tr-bbn} becomes severe at comparatively low $T_R$ (i.e., as it starts tending closer to $T_\text{BBN}$), as shown by the gray shaded region. Also, a larger $n$ requires a larger $g_X$ to produce the right relic for a fixed DM mass, as we already realized in Fig.~\ref{fig:dm-yld}.

\subsubsection*{Dark matter direct detection}\label{sec:si-dd}
In the present set-up the DM-nucleon scattering cross-section can be hugely amplified because it takes place via $t$-channel mediation of the light scalar $h_2$ (along with the SM-like Higgs $h_1$). The DM-nucleus scattering cross-section in this case has the form~\cite{Duch:2017khv}
\begin{equation}
\frac{d\sigma_{XN}}{dq^2}\left(q^2\right)=\frac{\sigma_{XN}}{4\mu_{XN}^2 v^2}\,F\left(q^2\right)\,,
\label{eq:ddx}
\end{equation}
where $v_\text{DM}$ is the DM velocity in the lab frame and the DM-nucleus reduced mass is given by $\mu_{XN}=m_X\,m_N/\left(m_X+m_N\right)$, and
\begin{equation}
\sigma_{XN} =\frac{\lhs^2\,f_N^2\,m_X^2\,m_N^2\,\mu_{XN}^2}{\pi\,m_{h_1}^4m_{h_2}^2\left(m_{h_2}^2+4\mu_{XN}^2\,v_\text{DM}^2\right)}\,,    
\end{equation}
is the total cross-section with the effective DM-nucleon coupling is $f_N\approx 0.3$~\cite{Cline:2013gha}. Here the factor $m_{h_2}^2+4\mu_{XN}^2 v_\text{DM}^2$ in the denominator represents the light mediator effects~\cite{Duch:2017khv,Duch:2019vjg}. Note that, the portal coupling $\lhs$ is not an independent parameter, but can be expressed in terms of $g_X,\,m_X$~\cite{Barman:2021lot}. The function $F(q^2)$ is defined as~\cite{Geng:2016uqt}
\begin{figure}[htb!]
    \centering
    \includegraphics[scale=0.34]{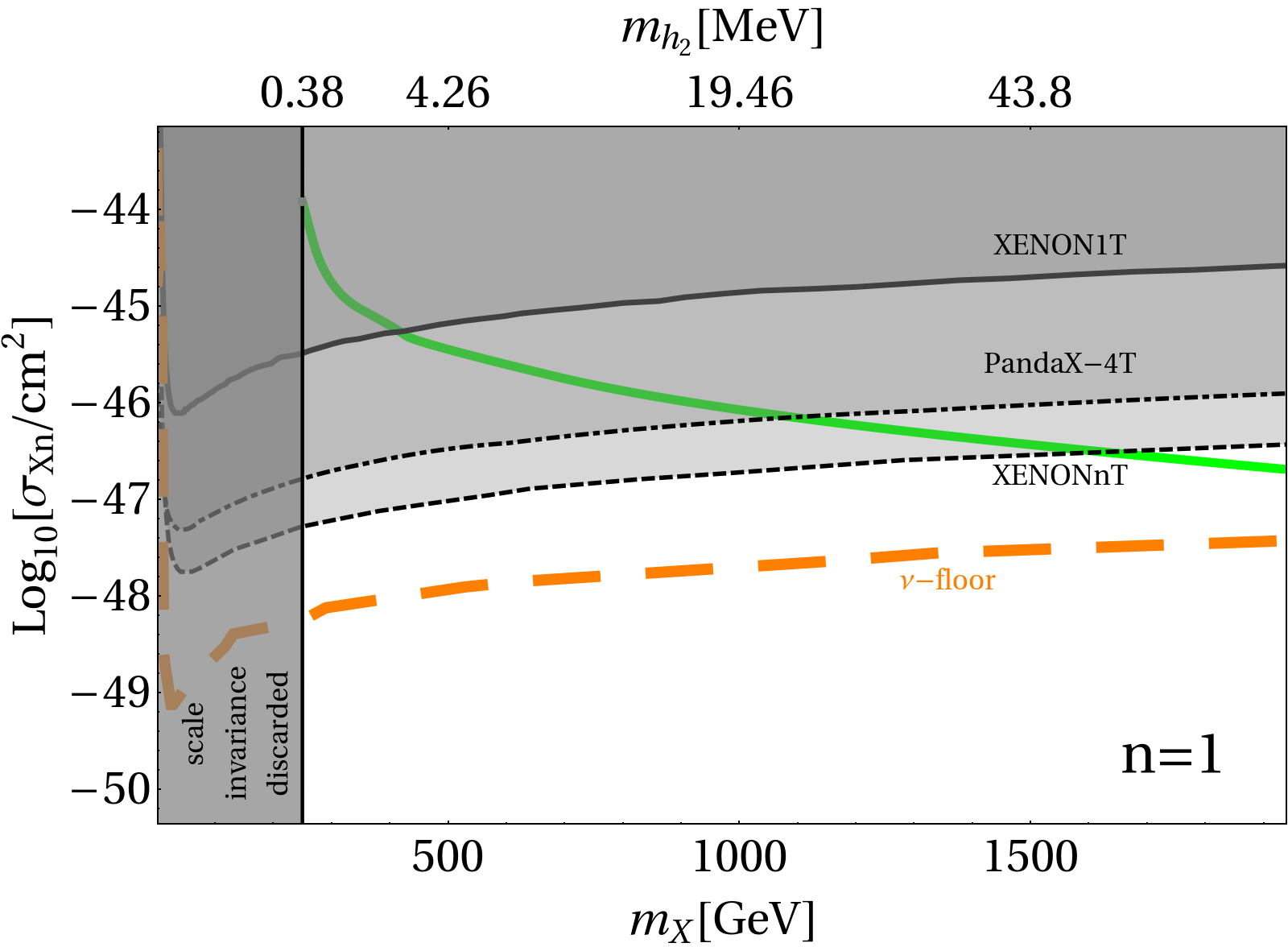}~~~~\includegraphics[scale=0.34]{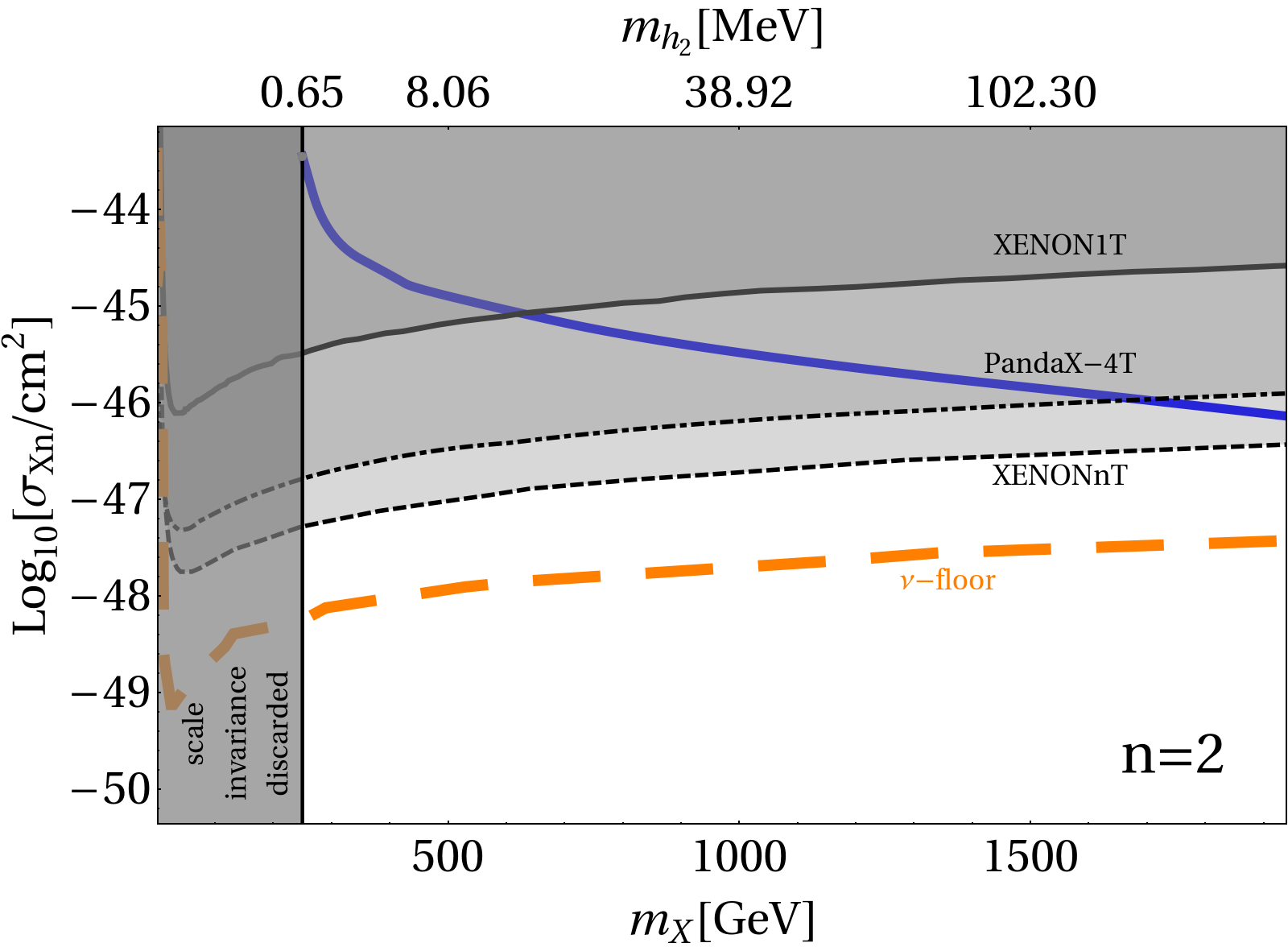}\\ \vspace{0.5cm}
    \includegraphics[scale=0.34]{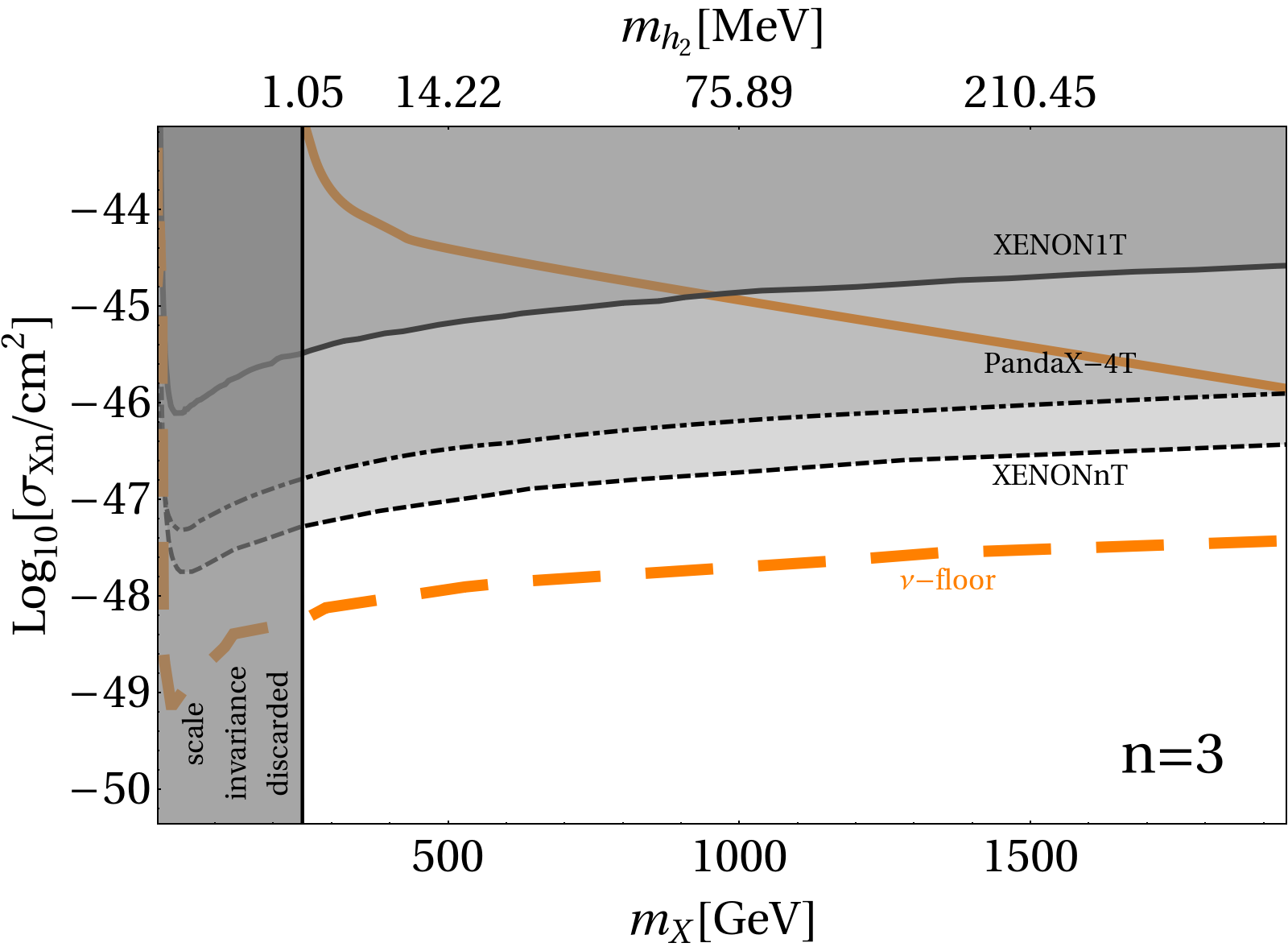}~~~~
    \includegraphics[scale=0.34]{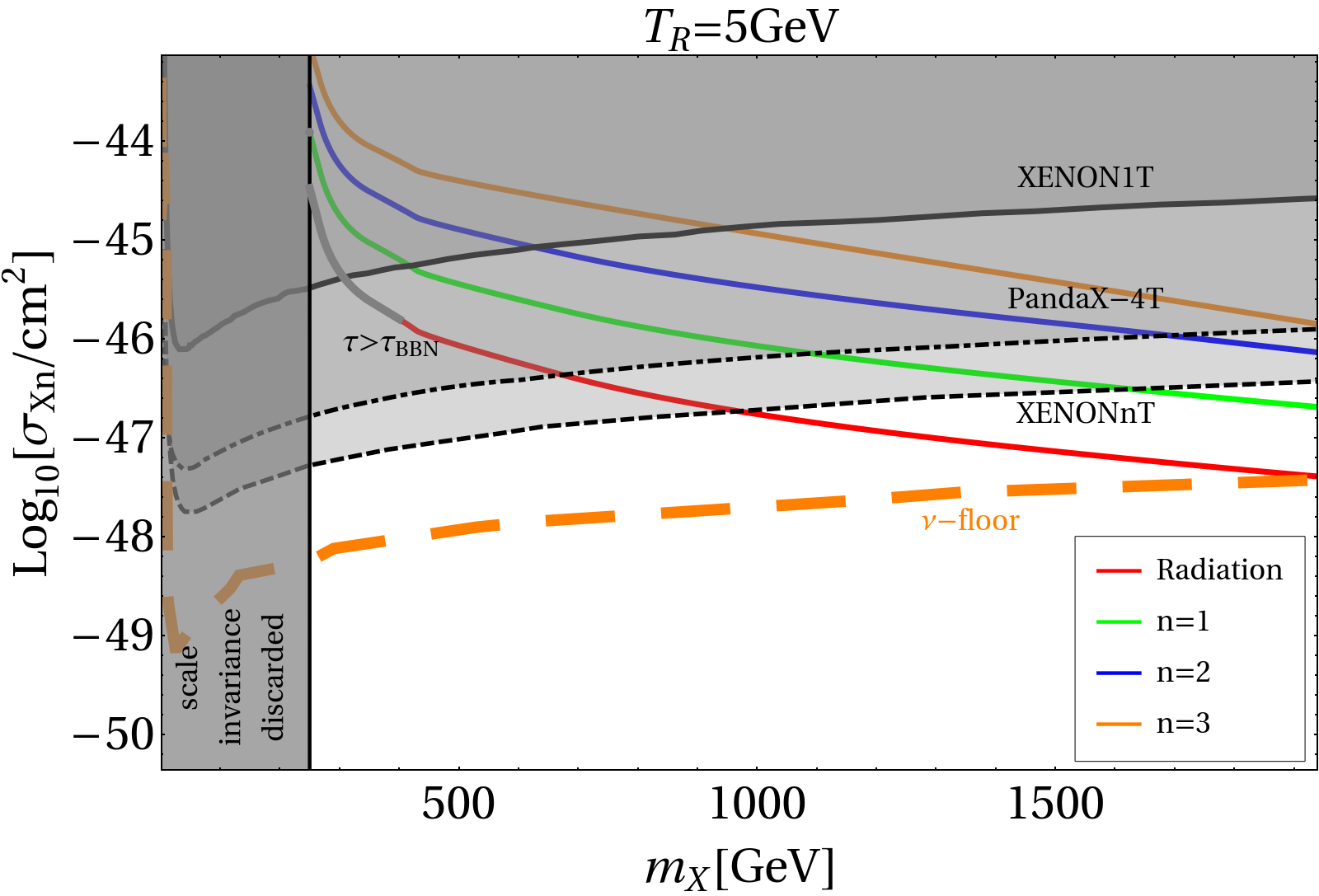}
    \caption{\it Top: Spin-independent direct detection constraint on the relic density allowed DM parameter space for $n=\{1,\,2\}$. Exclusion limits from XENON1T, XENONnT, PandaX-4T and the $\nu$-floor are shown, along with relic density allowed contours for different choices of $n$ by fixing $T_R=5$ GeV. In each case corresponding $h_2$ mass is also depicted along the top horizontal axis. The vertical gray shaded region in the left forbids $m_X<240$ GeV following Eq.~\eqref{eq:mh2}. Bottom: Same as top, but for $n=3$ (left) and a comparison with standard cosmology $(n=0)$ (right).}
    \label{fig:rel-dd}
\end{figure}
\begin{equation}
F(q^2) = \frac{\Bigl(1+q_\text{min}^2/m_{h_2}^2\Bigr)\Bigl(1+q_\text{ref}^2/m_{h_2}^2\Bigr)}{\Bigl(1+q^2/m_{h_2}^2\Bigr)^2}\,,
\end{equation}
which encodes the effects of the light mediator. Usually, $q_\text{min}^2$ is very small compared to the other scales appearing in the process, and thus can be taken to be zero, while  $q_\text{ref}^2=4\mu_{XN}^2v_\text{DM}^2$ is related to the energy thresholds of DM direct detection experiments. In the limit $m_{h_2}^2\gg q^2\sim 4\mu_{XN}^2v_\text{DM}^2$, the form factor $F(q^2)\approx 1$ and we recover the conventional DM-nucleon scattering cross-section for contact interactions. But for  $m_{h_2}^2\ll q^2$, Eq.~\eqref{eq:ddx} will have extra $q^2$ dependence characterized by $F(q^2)$. 
Because of the scale invariance of the theory, the DM-nucleus scattering cross-section turns out to be $\sigma_{XN}\propto g_X^4$, which implies large $g_X$ will result in large $\sigma_{XN}$, making the parameter space vulnerable to direct detection bounds\footnote{In deriving the direct search bounds we do not consider RGE flow of $g_X$ from UV scale down to nuclear physics energy scale since it remains almost fixed to a very small value (set by freeze-in).}. We consider exclusion limits from XENON1T~\cite{XENON:2018voc} (black solid curve), and projected bounds from PandaX-4T~\cite{PandaX:2018wtu} (black dotdashed) and XENONnT~\cite{XENON:2020kmp} (black dashed) experiments which provide upper limit on the DM-nucleon scattering cross-section at 90\% C.L. For DM mass $m_X\lesssim 390$ GeV, the BBN bound due to the non-standard Higgs decay becomes important for standard radiation domination $(n=0)$, while for larger $n$ this bound becomes irrelevant as we can understand from the right panel of Fig.~\ref{fig:gx-mx}. Since a faster than usual expansion requires a larger $g_X$ for obtaining the right abundance, therefore it also gives rise to a larger DM-nucleus scattering cross-section. We thus find that for $n=3$, DM mass up to $\sim 940$ GeV becomes excluded from current XENON1T limit. For the same reason, we see, larger $n$ results in heavier $h_2$ for a fixed DM mass (top horizontal axis in Fig.~\ref{fig:rel-dd}), following Eq.~\eqref{eq:mh2}. A heavier mediator is thus more constrained from direct search experiments, typically for DM mass $m_X\lesssim 2$ TeV, above which the DM-nucleon scattering loses its sensitivity. For $m_{h_2}\gtrsim\mathcal{O}(100\,\rm MeV)$ (equivalently, $m_X\gtrsim\mathcal{O}(50\,\rm TeV)$), limits from other experiments, like, beam dump or collider becomes more relevant. For a lower $T_R$ the direct search constraints become more stringent since in that case even larger coupling is needed to satisfy the relic density bound. In passing we would like to mention that since scale invariance forbids us to go to MeV-scale DM in the present model, hence direct detection bound from electron scattering experiments become irrelevant here. 

\section{Laboratory probe for Non-Standard Cosmology}
\label{sec:lab-probe}
We now investigate the prospects of searching the light scalar present in this theory in different intensity and lifetime frontier experiments and therefore probing the DM parameter space in turn, thanks to the underlying scale invarinace. Dark sectors with light degrees of freedom can be probed with a variety of experiments at the luminosity frontier, including proton~\cite{Batell:2009di, deNiverville:2011it, deNiverville:2012ij, Kahn:2014sra, LBNE:2013dhi, Soper:2014ska, Dobrescu:2014ita, Coloma:2015pih, deNiverville:2016rqh, MiniBooNE:2017nqe, MiniBooNEDM:2018cxm, MATHUSLA:2018bqv, FASER:2018bac}, electron~\cite{Bjorken:2009mm, Izaguirre:2013uxa, Diamond:2013oda, Izaguirre:2014dua, Batell:2014mga, BaBar:2017tiz, Berlin:2018bsc, Banerjee:2019pds} and positron fixed target facilities~\cite{Accardi:2020swt,Nardi:2018cxi}\footnote{For a very recent review on BSM searches in forward physics facilities see Ref.~\cite{Feng:2022inv}.}. These experiments are capable of probing extremely small mixing angles at great accuracy. In context with the present scenario, this will not only be a probe for DM, but also that of the cosmological history of the pre-BBN era of the Universe since the DM yield hugely depends on the fast expansion parameters $\{n,\,T_R\}$. Although light dark sector scenarios have been explored in great detail over the past decade~\cite{Bramante:2016yju, Alexander:2016aln, Battaglieri:2017aum, Darme:2017glc, Winkler:2018qyg, Egana-Ugrinovic:2019wzj, Okada:2019opp, Foroughi-Abari:2020gju, Agrawal:2021dbo,Nardi:2018cxi}, but as we already have established in~\cite{Barman:2021lot}, because of scale invariance, the resulting parameter space is extremely constrained and predictive for the present set-up. Together, here we establish how an alternative early Universe cosmology is capable of enhancing the detectability of freeze-in DM in potential forward search facilities. 
\begin{figure}[htb!]
    \centering
    \includegraphics[scale=0.37]{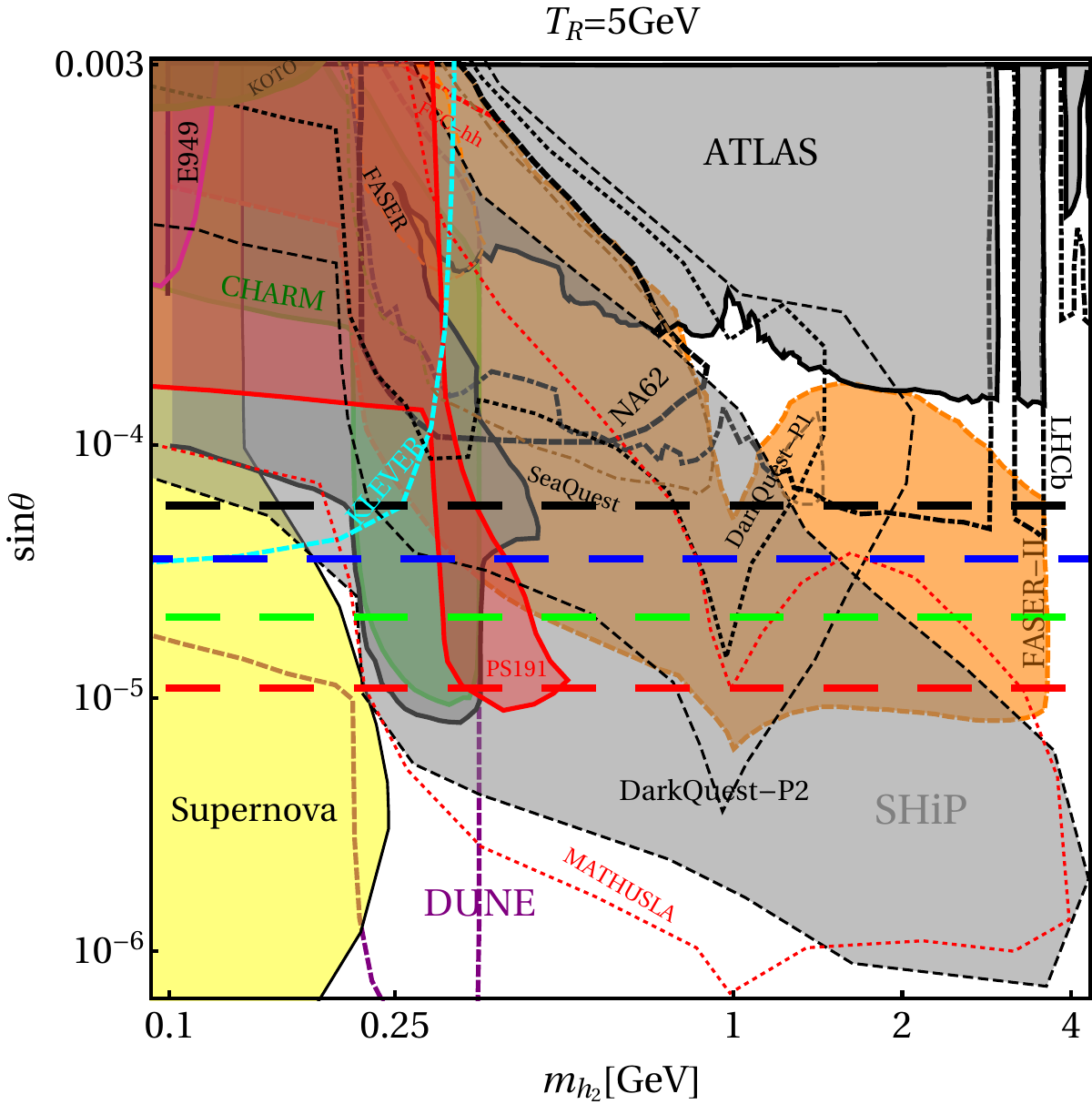}~~~~    \includegraphics[scale=0.37]{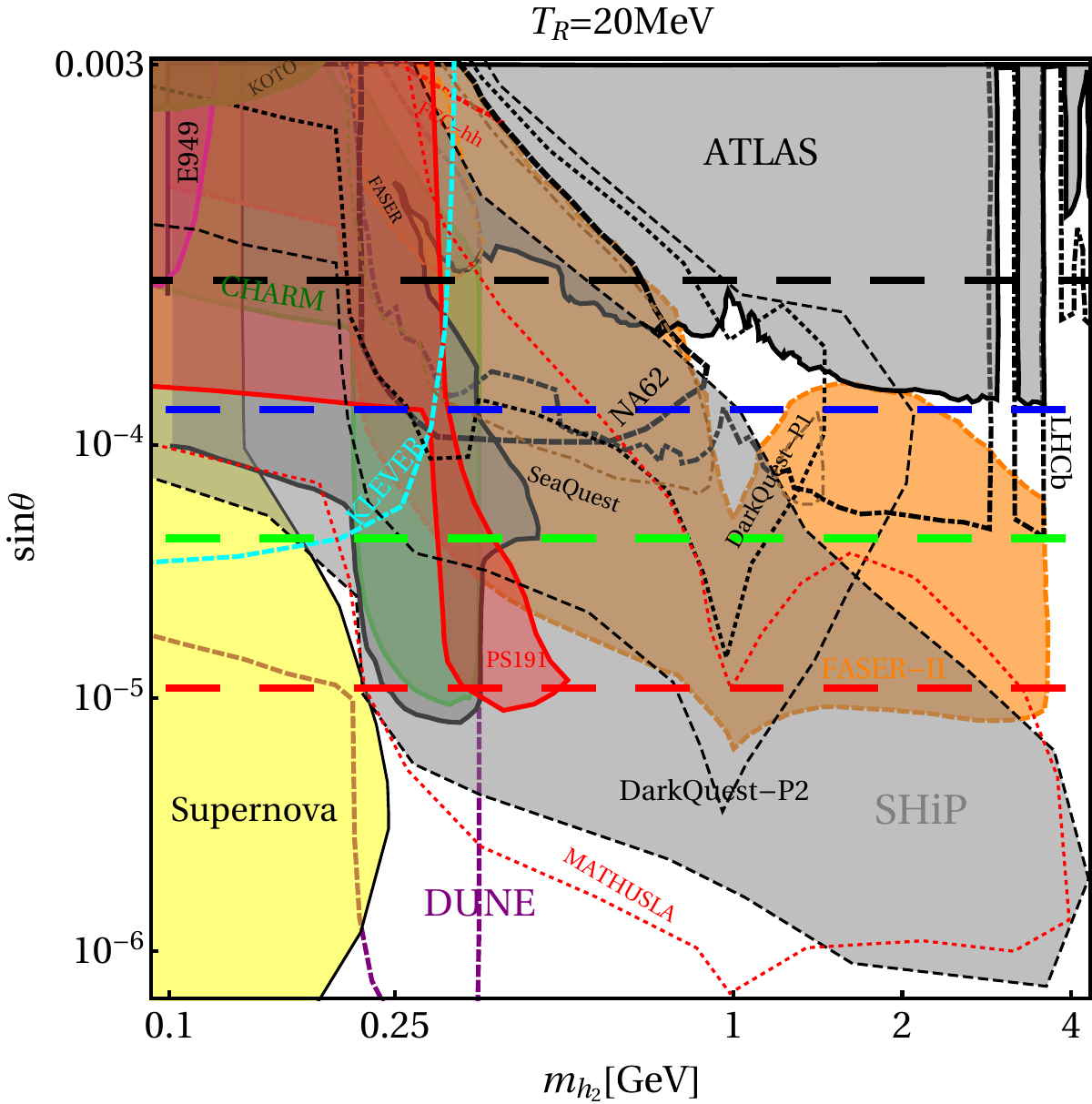}
    \caption{\it The thick straight dashed lines (parallel to the horizontal axis) with different colours show the DM parameter space complying with the PLANCK observed relic abundance and satisfying spin-independent direct search exclusion limit in $\sin\theta-m_{h_2}$ plane for $n=\{0,1,2,3\}$ from bottom to top with $T_R=5$ GeV (left) and $T_R=20$ MeV (right). Experimental limits are shown from E949~\cite{E949:2008btt}, CHARM~\cite{CHARM:1985anb}, NA62~\cite{Gonnella_2017}, FASER-I\&II~\cite{Feng:2017vli, FASER:2018eoc, FASER:2018bac, FASER:2019aik}, FCC-hh~\cite{FCC:2018vvp, DEV2017179}, ATLAS~\cite{ATLAS:2012qaq,ATLAS:2014jdv, Chalons:2016jeu, Robens:2015gla}, SeaQuest~\cite{Berlin:2018pwi}, LHCb~\cite{Gligorov:2017nwh}, KLEVER~\cite{Moulson:2018mlx}, DUNE~\cite{DUNE:2015lol, Berryman:2019dme}, DarkQuest-Phase2~\cite{Batell:2020vqn}, MATHUSLA~\cite{Curtin:2018mvb}, SHiP~\cite{SHiP:2015vad} and PS191~\cite{Bernardi:1985ny, Gorbunov:2021ccu}. All existing limits are solid lines, while future bounds are shown in broken lines.}
    \label{fig:expt-lim}
\end{figure}
The scale-invariance of the theory, along with the requirement of satisfying the right relic abundance, fixes $\sin\theta$ to a constant value, which we have dubbed as the {\it Scale Invariant FIMP Miracle} in~\cite{Barman:2021lot}. For a standard radiation dominated Universe, we found this value to be $\sim 10^{-5}$~\cite{Barman:2021lot}. In a non-standard scenario, with the increase in $n$, for a fixed $T_R$, one can make this mixing angle larger, still satisfying constraints from relic abundance and direct search. This is what we have shown in Fig.~\ref{fig:expt-lim}, where different coloured straight dashed lines, parallel to the horizontal axis, denote different choices of $n$ including the standard scenario shown in red. The improvement in the scalar mixing is easy to understand. On one hand, the underlying scale-invarinace determines the mixing through $\sin\theta\approx v_h\,g_X/m_X$, while on the other hand, a larger $n$ calls for a larger $g_X$ compared to the standard radiation domination, to satisfy the observed relic for a given $m_X$ (cf. Fig.~\ref{fig:gx-mx}). Here we see, with $T_R=20$ MeV (right panel), $n\gtrsim 2$ is already in tension from collider bounds due to ATLAS and LHCb. This implies, non-standard cosmological scenarios mimicking kination phase are in tension for the present framework for comparatively lower $T_R$. On the other hand, for $T_R=5$ GeV (left panel), this bound is much relaxed and $n\lesssim 4$ is allowed (although falls within the projected limts from ATLAS \& LHCb), implying there is still room for cosmological scenarios faster than quintessence for comparatively larger $T_R$. Note that, direct detection bound from XENON1T also allows $n=3$ for DM mass $m_X\gtrsim 1$ TeV for $T_R=5$ GeV (cf. Fig.~\ref{fig:rel-dd}), that as well falls within PandaX-4T and XENONnT future projections. Thus, we can constrain the fast expansion parameter $n$ from two orthogonal searches, namely, direct detection and intensity or energy frontiers, narrowing down the viable cosmological scenarios for freeze-in production of DM in this model. We see, different choices of $n$ lie well within the reach of projected sensitivities from experiments like MATHUSLA, SHiP, DUNE, FASER-II, DarkQuest etc that cover different ranges of $m_{h_2}$ depending on their respective reaches. This in turn also probes different ranges of $g_X$ and $m_X$ that provides correct DM relic. We therefore conclude that a faster expansion results in a larger mixing, making the present model easily falsifiable with already existing or improved bounds at several forward search and collider facilities. 

\section{Conclusions}
\label{sec:concl}
In summary, we investigated in a minimal scale-invariant framework, associated to radiative pathway of electroweak symmetry breaking, dark higgs-portal freeze-in vector dark matter (DM) in an early Universe suffering a non-standard cosmological expansion history due to $\varphi$-driven energy density. 
Parameterizing such a cosmological expansion history with modified Hubble background, we studied the consequences of DM freeze-in during the non-standard cosmological era, particularly in determining the coupling required to satisfy the observed relic density along with concrete predictions and constraints from direct detection and forward search facilities. Here we would like to emphasize that even though our conclusions are based on a scenario where the Universe expands faster than the standard radiation dominated background during $\varphi$-domination, but the same conclusion also holds for another well motivated alternative cosmological scenario, namely early matter domination (EMD) (for a review, see, for example~\cite{Allahverdi:2020bys}). In this case as well one needs to have a larger coupling (compared to standard radiation domination) with the visible sector to produce observed freeze-in DM abundance. There may be drastic modifications in the cosmic history due to several effects arising from the presence of this new species, however addressing such intricate details is beyond the scope of this paper and we concentrate only on the impact of this modified cosmological history on DM production. Considering our scenario, we find the following salient features that manifest

\begin{itemize}

    \item We showed non-standard cosmological era can be probed at experiments like FASER, DUNE, MATHUSLA, SHiP etc, typically for $n\lesssim 4$ with $T_R=5$ GeV to ensure non-thermal production of the DM (Fig.~\ref{fig:expt-lim}). The present model given is just for an example, but the prescription is applicable for any DM model with similar attributes and for portals other than the Higgs-portal. 
    
    \item As we previously showed in Ref.~\cite{Barman:2021lot}, scale-invariant FIMP Miracle dictated a single coupling ($\sin \theta \simeq 10^{-5}$) value for which one can satisfy the relic density (and also direct detection limits from XENON1T) fell within the reach of only a few of the low energy detectors. Here, in a non-standard cosmological set-up, with faster expansion history of the Universe, we found even with larger coupling the DM can remain out-of-equilibrium in early Universe (Fig.~\ref{fig:gx-mx}). We thus showed, non-thermal relic may get accumulated slowly from the SM bath (typical feature of IR freeze-in) with  larger mixing $\sin \theta \sim 10^{-3}$ that can may lie within the reach of experiments like ATLAS and LHCb, depending on the non-standard parameters $\{n,\,T_R\}$. Various $n\lesssim 4$ (which corresponds to different non-standard cosmologies) all satisfying the PLANCK observed relic density, can therefore be searched in ongoing/proposed experiments. We again remind the readers that this is a {\it direct} probe of the freeze-in coupling responsible for the right relic, courtesy to the scale-invariant nature of the model.
    
    \item We note that the model is extremely economical, given the fact that the only independent parameters are the new gauge coupling $g_X (\equiv\sin \theta)$, transition temperature $T_R$ and DM mass $m_X$, once a particular cosmology is fixed. But due to scale-invariance, one of them again gets fixed to satisfy the correct DM abundance, essentially leaving {\it two} free parameters which give us testable pre-BBN cosmology. However, for various $n$, BBN prevents us from going to very low $T_R$, following Eq.~\eqref{eq:tr-bbn} (Fig.~\ref{fig:ntr}).
    
    \item Finally, we showed such a probe of non-standard cosmological era in a scale-invariant set-up not only can be tested at the intensity frontiers, but due to the underlying symmetry there is a complementarity between direct detection (Fig.~\ref{fig:rel-dd}) versus lifetime frontier (Fig.~\ref{fig:expt-lim}). So, we will be able to observe signals at both the sets of otherwise unrelated experimental facilities, which should attract attention for more efforts in such directions of investigation in future.
    
\end{itemize}

Given such a spectacular relevance of scale invariance with respect to UV-completion of BSM theories, in context to the hierarchy problem, on one hand, while on the other hand, of the freeze-in mechanism of DM production in early Universe along with the availability of currently running and upcoming several light dark sector (intensity frontier and lifetime frontier) experiments, we showed that pre-BBN non-standard cosmology with FIMP DM candidate can be searched for in the laboratory including DM direct detection experiments with no other concrete probes of non-standard cosmological history except probably Primordial Gravitational Waves (GW) ($\Omega_{\rm GW} h^2$) at various GW detectors~\cite{Bernal:2020ywq,DEramo:2019tit} or dark radiation degrees of freedom ($N_{\rm eff}$)~\cite{Paul:2018njm}. This leads to predictive FIMP candidates with larger couplings together with non-standard pre-BBN history which we look to test in very near future. A combined analysis of low energy frontier, collider and direct detection laboratories along with GW and dark radiation observables in a freeze-in DM model with non-standard cosmology, along with complementary predictions: cosmology versus laboratory observables will be hugely interesting prospects, but is beyond the scope of the present manuscript and shall be discussed in a future work.

\section*{Acknowledgements}
The authors would like to thank Rouzbeh Allahverdi for comments, and Giorgio Arcadi and Francesco D'Eramo for providing feedback on the manuscript. BB received funding from the Patrimonio Autónomo - Fondo Nacional de Financiamiento para la Ciencia, la Tecnología y la Innovación Francisco José de Caldas (MinCiencias - Colombia) grant 80740-465-2020. This project has received funding /support from the European Union's Horizon 2020 research and innovation programme under the Marie Sklodowska-Curie grant agreement No 860881-HIDDeN. 

\appendix
\section{Annihilation cross-sections for freeze-in}\label{sec:app-ann}
Here we gather the analytical expressions for all the relevant 2-to-2 annihilation cross-section of the SM particles to DM final states. All SM leptons are denoted by $\ell$, quarks by $q$ and gauge bosons by $V\in W\,,Z$. The total decay width of the SM-like Higgs is given by $\Gamma_{h_1}\simeq 4$ MeV~\cite{CMS:2016dhk}.

\begin{equation}\begin{aligned}
&\sigma\left(s\right)_{\ell\ell\to XX}\simeq \frac{g_X^4\,m_\ell^2}{64\pi\,s}\,\frac{\sqrt{\left(s-4m_X^2\right)\,\left(s-4m_\ell^2\right)}}{\left(s-m_{h_1}^2\right)^2+\Gamma_{h_1}^2\,m_{h_1}^2}\Biggl(\frac{m_{h_1}^2-m_{h_2}^2}{s-m_{h_2}^2}\Biggr)^2\Biggl(\frac{s^2-4m_X^2s+12m_X^4}{\left(m_X^2+g_X^2v_h^2\right)^2}\Biggr)
\\&
\sigma\left(s\right)_{qq\to XX}\simeq\frac{g_X^4\,m_q^2}{192\pi\,s}\frac{\sqrt{\left(s-4m_X^2\right)\,\left(s-4m_q^2\right)}}{\left(s-m_{h_1}^2\right)^2+\Gamma_{h_1}^2\,m_{h_1}^2}\Biggl(\frac{m_{h_1}^2-m_{h_2}^2}{s-m_{h_2}^2}\Biggr)^2\Biggl(\frac{s^2-4m_X^2s+12m_X^4}{\left(m_X^2+g_X^2v_h^2\right)^2}\Biggr)
\\&
\sigma\left(s\right)_{VV\to XX}\simeq\frac{g_X^4}{288\pi\,s}\,\sqrt{\frac{s-4m_X^2}{s-4m_V^2}}\,\Biggl(\frac{m_{h_1}^2-m_{h_2}^2}{s-m_{h_2}^2}\Biggr)^2\,\Biggl(\frac{1}{\left(s-m_{h_1}^2\right)^2+\Gamma_{h_1}^2\,m_{h_1}^2}\Biggr)\\&\Biggl[\frac{\left(s^2-4m_X^2s+12m_X^4\right)\left(s^2-4m_V^2s+12m_V^4\right)}{\left(m_X^2+g_X^2v_h^2\right)^2}\Biggr]
\\&
\sigma\left(s\right)_{h_1 h_1\to XX}\simeq \frac{g_X^4}{32\pi s}\Biggl(\frac{m_{h_1}}{m_X}\Biggr)^4 \frac{\left(s+2m_{h_1}^2\right)^2}{\left(s-m_{h_1}^2\right)^2+\Gamma_{h_1}^2\,m_{h_1}^2}\,\sqrt{\frac{s-4m_X^2}{s-4m_{h_1}^2}}\,\Biggl(1-\frac{4m_X^2}{s}+\frac{12m_X^4}{s^2}\Biggr).
\end{aligned}    
\end{equation}

\noindent The last expression is derived by keeping only the leading order in the double expansion of $g_X$ and $m_{h_2}^2/m_X^2$.

\section{Decays of $h_2$}
\label{sec:app-h2-decay}

The partial decay widths to the SM final states are given by~\cite{Djouadi:2005gi,Krnjaic:2015mbs}

\begin{equation}\begin{aligned}
&\Gamma_{ff} = \frac{G_F\sin^2\theta\,N_c}{4\sqrt{2}}m_f^2 m_{h_2}\Biggl(1-\frac{4m_f^2}{m_{h_2}^2}\Biggr)^{3/2}
\\&
\Gamma_{\gamma\gamma}=\frac{G_F\sin^2\theta}{128\sqrt{2}}\frac{\alpha^2 m_{h_2}^3}{\pi^3}\Biggl|\sum_f N_c Q_f^2 \mathcal{A}_{1/2}\left(x_f\right)+\mathcal{A}_1\left(x_f\right)\Biggr|^2
\\&
\Gamma_{gg} = \frac{G_F\sin^2\theta}{36\sqrt{2}}\frac{\alpha_s m_{h_2}^3}{\pi^3}\Biggl|\frac{3}{4}\sum_q\mathcal{A}_{1/2}\left(x_q\right)\Biggr|^2
\end{aligned}\label{eq:decayW}
\end{equation}

and

\begin{equation}\begin{aligned}
&\mathcal{A}_{1/2}\left(x\right)=2\Bigl[x+\left(x-1\right)f(x)\Bigr]x^{-2}
\\&
\mathcal{A}_1(x)=-\Bigl[2x^2+3x+3\left(2x-1\right)f(x)\Bigr]x^{-2}
\end{aligned}
\end{equation}

with

\begin{equation}
f(x) \equiv
    \begin{cases}
        \text{Arc}\sin^2\sqrt{x} & x\leq 1,\\[8pt]
        -\frac{1}{4}\Bigl[\log\frac{1+\sqrt{1-x^{-1}}}{1-\sqrt{1-x^{-1}}}-i\pi\Bigr]^2 & x>1\,.
    \end{cases}
\end{equation}

\noindent where $x_i=m_{h_2}^2/4m_i^2$, $N_c$ is the number of colors for a given fermion species and $Q_f$ is its electromagnetic charge. 

\section{BBN constraint}\label{sec:bbn}

The effect of the new species $\varphi$ can be parametrized by an effective number of relativistic degrees of freedom (DOF)

\begin{equation}\begin{aligned}
& \rho\left(T\right) = \frac{\pi^2}{30}g_{\star\text{eff}} T^4     \end{aligned}\end{equation}

with 

\begin{equation}\begin{aligned}
& g_{\star\text{eff}} = g_\star^\text{SM}+\Delta g_\star^\varphi=\Bigl(2+\frac{7}{8}\times 4\Bigr)+\Bigl(2+\frac{7}{8}\times N_\nu\Bigr)\,  
\end{aligned}
\end{equation}

\noindent where the effect of $\varphi$ is parametrized by taking into acount total number of effective neutrinos $N_\nu=N_\nu^\text{SM}+\Delta N_\nu$ with $N_\nu=3.044$~\cite{Akita:2020szl,Froustey:2020mcq,Bennett:2020zkv} for the SM. Thus, the temperature dependence of the number of additional neutrino species can be expressed as

\begin{equation}\begin{aligned}
& \Delta N_\nu  = \frac{4}{7}\, g_\star\left(T_R\right)\,\Biggl(\frac{g_{\star s\left(T\right)}}{g_{\star s}\left(T_R\right)}\Biggr)^{(4+n)/3}\,\Biggl(\frac{T}{T_R}\Biggr)^n.   
\end{aligned}
\end{equation} 

For $T_R\sim T_\text{BBN}$ we have $g_\star\left(T_R\right)=\frac{43}{4}$, which results

\begin{equation}\begin{aligned}
& \Delta N_\nu\simeq 6.14\Biggl (\frac{T}{T_R}\Biggr)^n.    
\end{aligned}
\end{equation}

Following~\cite{Cyburt:2015mya} we find $2.3\leq\Delta N_\nu\leq 3.4$ at 95\% CL, and considering $T\simeq 1~\rm MeV$ we obtain, in order to satisfy the BBN bounds we should have $T_R \gtrsim \left(15.4\right)^{1/n}~\text{MeV}.$

\bibliographystyle{JHEP}
\bibliography{Bibliography}

\end{document}